\documentclass[reprint,aps,onecolumn,floatfix,nofootinbib,nobibnotes]{revtex4-2}
\usepackage[a4paper,margin=1in]{geometry}  
\usepackage{amsmath, amssymb, graphicx, color} 
\usepackage{hyperref}  
\usepackage{graphicx}
\usepackage{dcolumn}
\usepackage{bm}
\usepackage{leftidx}
\usepackage{txfonts}
\usepackage{physics}
 
\usepackage{enumitem}
\usepackage{appendix}
\usepackage{bbm} 

\begin{document}
\title{Foundations of Wilson lines and loops from basic concepts to gauge field reconstruction}

 \author{Shokoufe Faraji}
  \email{s3faraji@uwaterloo.ca}

  \affiliation{Department of Physics and Astronomy, University of Waterloo, Canada}
  \affiliation{Perimeter Institute for Theoretical Physics, Canada}
 \affiliation{Center of Applied Space Technology and Microgravity (ZARM), University of Bremen, Germany}




\begin{abstract}
This review presents a concise and unified treatment of Wilson lines and Wilson loops in continuum gauge theory, emphasizing their interpretation as parallel transport and their status as gauge covariant transporters and gauge invariant loop observables. Starting from the problem of comparing charged fields at separated points, we derive the path ordered exponential formulation and collect the key kinematical identities: gauge covariance, composition under path concatenation, inversion, and base point dependence. We then explain how local information is encoded in suitable small segment and small loop limits, and how these limits enter reconstruction questions for the underlying gauge connection. Seen as an inverse problem, this perspective clarifies how much of the gauge connection is encoded in loop observables and why traced Wilson loops, while gauge invariant, are intrinsically lossy without additional input. Special attention is given to clarifying common sources of confusion, in particular the distinction between untraced holonomies and traced Wilson loops, and the role of reference path choices when relating open transports to based loop data. The presentation is intended to serve both as an entry point for newcomers and as a compact reference for readers in high energy theory, mathematical physics, and related fields. 

\medskip

\end{abstract}

\maketitle
\tableofcontents  

\section{Introduction}

Gauge theories describe interactions in terms of a gauge connection $A_\mu(x)$ taking values in the Lie algebra of a gauge group $G$.
While local quantities such as the field strength encode the curvature of the connection, many central questions (including confinement diagnostics, topological effects, and nonperturbative structure) are more naturally phrased in terms of nonlocal parallel transport and loop observables.
From a differential geometric viewpoint, these observables are built from the parallel transport defined by a connection on a principal bundle (see e.g.,\ \cite{KobayashiNomizu:1963fdg1}).
Given a path $\gamma$ from $y$ to $x$, the associated parallel transporter (holonomy) $U(\gamma)$ compares charged fields at separated points in a gauge covariant way.
For a closed loop $C$ based at $x$, the holonomy $U(C)$ transforms by conjugation under a gauge transformation, and its trace in a representation $R$ defines the gauge invariant Wilson loop $W_R(C)=\mathrm{Tr}_R\,U(C)$.
A recurring theme throughout this review is the distinction between holonomy level and trace level statements: taking a trace produces a manifestly gauge invariant observable but simultaneously discards information.
For standard perspectives on Wilson loops and related nonlocal observables, see, e.g.,\ \cite{Polyakov:1987gfs,Makeenko:2002mcgt}.

Wilson loops were introduced by Wilson in the 1970s in his lattice formulation of non-Abelian gauge theories, where they provide a natural gauge invariant probe of confinement in pure gauge theory \cite{Wilson:1974sk}. In that setting, group elements are assigned to oriented lattice links, and their ordered product around a closed path gives the corresponding loop holonomy. The expectation value of a large Wilson loop can exhibit characteristic area law or perimeter law behavior; in the presence of dynamical matter in the fundamental representation, screening modifies the strict area law criterion.


Since Wilson's original work, Wilson lines and loops have become central tools across high-energy theory, mathematical physics, and condensed matter.
In lattice gauge theory they remain standard gauge invariant probes of static potentials and string tensions (see, e.g.,\ \cite{Rothe:2005lattice,Greensite:2011confinement}).
In gauge/string duality, Wilson loops are related to string worldsheets ending on the loop contour and, at strong coupling, to minimal surface problems in the dual geometry \cite{Maldacena:1998im,Rey:1998ik,Drukker:1999adh,DrukkerGrossOoguri:1999}.

In supersymmetric gauge theories, BPS Wilson loops provide protected observables that can in certain cases be evaluated exactly using matrix model or localization methods \cite{EricksonSemenoffZarembo:2000,DrukkerGross:2000,Pestun:2007}. In Chern-Simons theory, Wilson loop expectation values give rise to knot and link invariants, most famously through Witten's relation between Chern-Simons theory and the Jones polynomial \cite{Witten:1988hf,Bruegmann:1994zr,FrChS:1989}.
Holonomies also furnish basic background-independent variables in loop based approaches to quantum gravity \cite{Rovelli:1989jn,Gambini:1996ph,Ashtekar:2004eh,Thiemann:2007zz}, while related topological constructions underlie effective descriptions of phases such as fractional quantum Hall states \cite{Wen:1992uk,Ardonne:2003wa,Fradkin:2013fqh,Tong:2016kpv}. At the dynamical level, Wilson loops organize loop equation approaches to gauge theory \cite{MakeenkoMigdal:1981loops,Makeenko:2002mcgt}.
At the quantum operator level, they also possess nontrivial renormalization properties: contour singularities such as cusps and self-intersections introduce additional anomalous dimensions and operator mixing \cite{BrandtNeriSato:1981,KorchemskyRadyushkin:1987,BerweinBrambillaVairo:2013}.
These developments are important parts of the broader Wilson loop literature; the principal focus of the present review, however, is the underlying kinematics, geometry, and reconstruction problem.


The aim of this review is to give a self-contained and mathematically careful account of Wilson lines and Wilson loops in continuum gauge theory, emphasizing the structure underlying parallel transport, gauge covariance, base point dependence, and the information contained in appropriate infinitesimal limits. Although the individual ingredients are standard, they are often presented separately across the gauge theory, differential geometric, and loop variable literature. Here they are organized around a single inverse question: which loop or holonomy data are sufficient to recover the underlying gauge connection up to gauge, and what information is lost when one passes from holonomies to traced Wilson loops?

This inverse viewpoint is both conceptually sharp and practically useful. At the holonomy level, sufficiently regular parallel transport data can determine a local connection up to gauge on a patch.
At the traced level, one has only character data: just as the trace of a matrix does not determine the matrix itself, a traced Wilson loop does not in general determine the corresponding holonomy. Any reconstruction from traced data therefore requires additional input, such as richer representation content, untraced holonomies, controlled open path data, or further admissibility conditions.
Early path-dependent formulations and reconstruction analyses include \cite{Mandelstam:1962qed,Giles:1981reconstruct,Barrett:1991holonomy}.

Concretely, we first construct Wilson lines from the requirement of comparing charged fields at separated points and explain how their local expressions are supplemented by transition functions on nontrivial bundles.
We then collect the structural identities of path ordered transport, analyze the short segment and small loop limits that recover the connection and curvature, and explain the roles of base point transport and reference paths.
Finally, we examine the inverse problem at both the holonomy and trace levels, including the algebraic, positivity, and regularity conditions that constrain admissible traced loop data.

\subsection*{Scope and conventions}

We work with a general gauge group $G$ and allow an arbitrary choice of representation $R$ when traces are taken. Gauge transformations are denoted by $g(x)$, while $U(\gamma)$ denotes parallel transport along a path $\gamma$; when necessary, $U_R(\gamma)$ denotes its matrix in the representation $R$.
Open Wilson lines are gauge covariant through their endpoint transformation law, whereas traced Wilson loops
$W_R(C)=\mathrm{Tr}_R\,U_R(C)$ are gauge invariant.
We consistently distinguish the holonomy around a closed loop from its traced Wilson loop and keep holonomy level statements separate from trace level statements.
Local reconstruction formulae are written in a chosen trivialization, while the transition function data required for a global definition on nontrivial bundles are discussed separately.

The paper is organized as follows.
Section~\ref{sec:first} introduces Wilson lines as parallel transport and discusses their local and globally patched definitions.
Section~\ref{sec:structure} develops their structural properties and infinitesimal limits.
Section~\ref{sec:constr} reconstructs finite parallel transport from infinitesimal segments.
Section~\ref{sec:covariant} discusses base point transport and covariant derivatives of loop holonomies.
Section~\ref{sec:recover} explains how open holonomies can be recovered from based loop holonomies using reference paths.
Section~\ref{sec:reconst} examines gauge field reconstruction from holonomy and traced loop data.
Section~\ref{sec:sum} summarizes the principal conclusions and broader implications.

\section{Wilson lines and Wilson loops}\label{sec:first}

\subsection{Comparing charged fields at separated points}

Let $M$ be spacetime, or more generally the manifold on which the gauge theory is defined, and let $G$ be the gauge group.
In a local trivialization, a matter field $\psi(x)$ transforming in a representation of $G$ transforms pointwise under
a gauge transformation $g:M\to G$ as
\begin{equation}
\psi(x)\ \longmapsto\ \psi'(x)=g(x)\,\psi(x),
\end{equation}
where $g(x)$ denotes the action of the gauge transformation in the chosen representation.
Because $g(x)$ varies with position, the transformations acting at $x$ and $y$ need not coincide. This prevents a direct gauge covariant comparison of field values at distinct points. For example, $\psi(x)-\psi(y)$ transforms as $g(x)\psi(x)-g(y)\psi(y)$ and therefore does not transform with a single group element.
The underlying reason is that gauge symmetry supplies no preferred identification of the internal degrees of freedom
at different points; such an identification must instead be provided by parallel transport. Geometrically, $\psi(x)$ and $\psi(y)$ lie in different fibers of the associated bundle, so there is no canonical identification between them
(see, e.g.,\ \cite{KobayashiNomizu:1963fdg1}).

To make such comparisons meaningful, one needs an additional structure: connection (gauge potential), which specifies how internal
degrees of freedom are transported along a chosen curve.
Given a path $\gamma:y\to x$, the connection assigns a parallel transporter $U(\gamma)$ such that
$U(\gamma)\psi(y)$ can be compared directly with $\psi(x)$.
The dependence on $\gamma$ is essential: in general, different paths between the same endpoints produce different
transporters, with the local path dependence governed by the curvature. Gauge covariance requires the endpoint transformation law
\begin{equation}
U(\gamma)\ \longmapsto\ U'(\gamma)=g(x)\,U(\gamma)\,g(y)^{-1}.
\end{equation}
Consequently, both $\psi(x)$ and $U(\gamma)\psi(y)$ transform with the same factor $g(x)$, and their difference
$\psi(x)-U(\gamma)\psi(y)$ is gauge covariant at the common endpoint $x$.
When a field in the dual representation is included, the corresponding bilocal quantity $\bar{\psi}(x)\,U(\gamma)\,\psi(y)$ is gauge invariant. The Wilson line is precisely this path dependent parallel transporter. In the next subsection we construct $U(\gamma)$ explicitly from the gauge potential, obtain its path ordered exponential representation, and establish its endpoint covariance.



\subsection{Wilson line as parallel transport}

Let $\gamma:[0,1]\to M$ be a smooth path with $\gamma(0)=y$ and $\gamma(1)=x$.
In a chosen local trivialization and representation, let write the connection one form as
$A=A_\mu(x)\,dx^\mu$. Its contraction with the tangent to the path is
\begin{equation}
A(\sigma)\ \equiv\ \dot{\gamma}^\mu(\sigma)\,A_\mu(\gamma(\sigma)).
\end{equation}
The parallel transporter from the initial point $y$ to the intermediate point $\gamma(\sigma)$ is denoted by
$U_\gamma(\sigma)\equiv U(\gamma;\sigma,0)$ and is defined as the unique solution of the initial value problem
\begin{align}\label{eq:parallel_transport_ode}
\frac{d}{d\sigma}U_\gamma(\sigma)
&= i\,A(\sigma)\,U_\gamma(\sigma),\\
U_\gamma(0)&=\mathbbm{1},
\end{align}
where $\mathbbm{1}$ denotes the identity matrix (identity operator) in the chosen representation. The Wilson line from $y$ to $x$ along $\gamma$ is then
\begin{equation}
U(\gamma)\equiv U_\gamma(1).
\end{equation}
More generally, parallel transport between two intermediate points on the same path is
\begin{equation}
U(\gamma;\sigma_2,\sigma_1)=U_\gamma(\sigma_2)\,U_\gamma(\sigma_1)^{-1}.
\end{equation}
Note that with the convention $D_\mu=\partial_\mu-iA_\mu$, the sign in Equation \eqref{eq:parallel_transport_ode} is fixed as written; when $A_\mu$ is Hermitian, the resulting transporter is unitary in a unitary representation. Integrating Equation \eqref{eq:parallel_transport_ode} gives
\begin{equation}\label{eq:volterra}
U_\gamma(\sigma)=\mathbbm{1}+ i\int_0^\sigma d\sigma_1\,
A(\sigma_1)\,U_\gamma(\sigma_1).
\end{equation}
Iterating \eqref{eq:volterra} generates the Dyson series
\begin{align}\label{eq:dyson}
U(\gamma)
&=\mathbbm{1}+\sum_{n=1}^{\infty} i^n
\int_{0\le \sigma_n\le\cdots\le\sigma_1\le 1} d\sigma_1\cdots d\sigma_n\;
A(\sigma_1)\cdots A(\sigma_n)\nonumber\\
&=\mathbbm{1}
+ i\int_0^1 d\sigma_1\,A(\sigma_1)
+ i^2\int_0^1 d\sigma_1\int_0^{\sigma_1} d\sigma_2\,A(\sigma_1)A(\sigma_2)+\cdots.
\end{align}
The series is conventionally summarized by the path ordered exponential
\begin{align}\label{eq:path_ordered_exp}
U(\gamma)
&=\mathcal{P}\exp\left(i\int_0^1 d\sigma\;A(\sigma)\right)\nonumber\\
&=\mathcal{P}\exp\left( i\int_0^1 d\sigma\;\dot{\gamma}^\mu(\sigma)\,A_\mu(\gamma(\sigma))
\right).
\end{align}
Here $\mathcal{P}$ places factors with larger parameter value $\sigma$ to the left.
Thus, $\mathcal{P}\exp$ is shorthand for the unique solution of the transport equation, rather than an ordinary
matrix exponential. If $A(\sigma_1)$ and $A(\sigma_2)$ commute for all parameter values, the ordering may be omitted; this is automatically the case in an abelian theory, whereas in a non-Abelian theory path ordering is generally
essential (see, e.g.,\ \cite{Polyakov:1987gfs,Makeenko:2002mcgt}).

\subsection{Local trivializations and global patching}
\label{subsec:global_patching}

The preceding formulas were written in a single local trivialization. This is sufficient for the local reconstruction questions considered in this review. On a topologically nontrivial gauge bundle, however, a single gauge potential need not be defined globally. The connection must then be described by local gauge potentials on overlapping patches, and the corresponding transition functions must be included whenever a Wilson line crosses from one patch to another. This issue already appears in the abelian theory, as in the Wu-Yang description of the Dirac monopole and in the cohomological formulation of topological quantization
\cite{WuYang:1976DiracStrings,Alvarez:1985tqc}.

For clarity, consider a $U(1)$ connection described on two patches $U_\alpha$ and $U_\beta$ by local one-forms
$A_\alpha$ and $A_\beta$. On the overlap, let the local descriptions of a charged field be related by
\begin{equation}
\psi_\alpha=e^{i\lambda_{\alpha\beta}}\psi_\beta .
\end{equation}
Compatibility with the convention $D=d-iA$ then requires
\begin{equation}\label{eq:abelian_patch_connection}
A_\alpha=A_\beta+d\lambda_{\alpha\beta}
\qquad \text{on }U_\alpha\cap U_\beta .
\end{equation}
Suppose that a path begins at $I\in U_\beta$, ends at $F\in U_\alpha$, and crosses the overlap at
$P\in U_\alpha\cap U_\beta$. The transporter from the $\beta$-description at $I$ to the
$\alpha$-description at $F$ is
\begin{align}\label{eq:patched_abelian_transport}
U_{\alpha\beta}(F,I)
&=\exp\,\left(i\int_P^F A_\alpha\right)e^{i\lambda_{\alpha\beta}(P)}
\exp\,\left(i\int_I^P A_\beta\right)\nonumber\\
&=\exp\,\left[ i\left( \int_I^PA_\beta+\lambda_{\alpha\beta}(P)+\int_P^F A_\alpha\right)\right].
\end{align}
The result is independent of the chosen crossing point $P$: moving $P$ within the overlap changes the two line integrals
by an amount that is cancelled by the corresponding change in $\lambda_{\alpha\beta}(P)$, as follows from
\eqref{eq:abelian_patch_connection}. For a path crossing several patches, the appropriate transition factor is inserted at every crossing.

The construction has the same form for a non-Abelian gauge group. If $\psi_\alpha=g_{\alpha\beta}\psi_\beta$ on an overlap, then
\begin{equation}
A_\alpha=g_{\alpha\beta}A_\beta g_{\alpha\beta}^{-1}
-i\,(d g_{\alpha\beta})g_{\alpha\beta}^{-1},
\end{equation}
and a crossing from $U_\beta$ to $U_\alpha$ contributes the transition matrix in the ordered product,
\begin{equation}
U_{\alpha\beta}(F,I)=U_\alpha(F,P)\,g_{\alpha\beta}(P)\,U_\beta(P,I).
\end{equation}
Thus transition matrices enter in the same path order as the local transporters. On triple overlaps the transition functions satisfy the cocycle condition
\begin{equation}
g_{\alpha\beta}\,g_{\beta\gamma}\,g_{\gamma\alpha}=\mathbbm{1}.
\end{equation}
For $U(1)$, with $g_{\alpha\beta}=e^{i\lambda_{\alpha\beta}}$, this becomes
\begin{equation}
\lambda_{\alpha\beta} +\lambda_{\beta\gamma}
+\lambda_{\gamma\alpha} = 2\pi n_{\alpha\beta\gamma},
\qquad n_{\alpha\beta\gamma}\in\mathbb{Z}.
\end{equation}
These cocycle data encode the global topological information of the gauge bundle. The patched expression above defines an open transporter with the appropriate endpoint covariance; it becomes gauge invariant only after the endpoints are contracted with charged fields or the path is closed and a class function\footnote{A class function is any function $f:G \to \mathbb{C}$ that is constant under conjugation: $f(hgh^{-1})=f(g)$.}, such as the trace, is taken. In this review we work locally in a chosen trivialization, so the infinitesimal reconstruction formulas are unaffected by this global layer, while the transition functions supply the additional information needed for a fully global description.

\subsection{Gauge transformations and Wilson loops}

Under a local gauge transformation $g(x)$, the gauge potential transforms (details in App.~\ref{app:gauge_trans_A}) as
\begin{equation}\label{eq:gauge_trans_A}
\begin{aligned}
A_\mu(x)\ \longmapsto\ A_\mu^{\,g}(x)
&=g(x)\,A_\mu(x)\,g(x)^{-1}
-i\,\big(\partial_\mu g(x)\big)\,g(x)^{-1}\\
&=g(x)\,A_\mu(x)\,g(x)^{-1}
+i\,g(x)\,\partial_\mu g(x)^{-1}.
\end{aligned}
\end{equation}
With the convention $D_\mu=\partial_\mu-iA_\mu$, this is precisely the transformation law that ensures
\begin{equation}
D_\mu^{\,g}\psi^{\,g}(x)=g(x)\,D_\mu\psi(x).
\end{equation}
The corresponding parallel transporter transforms covariantly at its endpoints. For
$\gamma:[0,1]\to M$ one has

\begin{equation}\label{eq:gauge_trans_U}
U(\gamma;\sigma_2,\sigma_1)\ \longmapsto\ U^{\,g}(\gamma;\sigma_2,\sigma_1)=g(\gamma(\sigma_2))\,U(\gamma;\sigma_2,\sigma_1)\,g(\gamma(\sigma_1))^{-1}.
\end{equation}
This follows directly from the transport equation: the right hand side of Equation \eqref{eq:gauge_trans_U} satisfies the transformed differential equation and the corresponding initial condition. In particular, for $\gamma:y\to x$ this reduces to the endpoint law
\begin{equation}
U(\gamma)\ \longmapsto\ U^{\,g}(\gamma)=g(x)\,U(\gamma)\,g(y)^{-1}.
\end{equation}
Thus an open Wilson line is gauge covariant rather than gauge invariant. A gauge invariant quantity can be formed by
including matter fields at both endpoints. For example, if
$\bar\psi(x)\mapsto\bar\psi(x)g(x)^{-1}$, then
\begin{equation}
\bar\psi(x)\,U(\gamma)\,\psi(y)
\end{equation}
is gauge invariant. Closing the path removes the distinction between the two endpoints but does not by itself make the untraced holonomy gauge invariant. If $C$ is a closed loop based at $x_0$ (so $C:x_0\to x_0$), its holonomy transforms by conjugation,

\begin{equation}
U(C)\ \longmapsto\ U^{\,g}(C)=g(x_0)\,U(C)\,g(x_0)^{-1}.
\end{equation}
Consequently any class function of $U(C)$ is gauge invariant. In particular, for a chosen representation $R$ of $G$,
the traced Wilson loop is the gauge invariant observable
\begin{equation}\label{eq:wilson_loop_def}
W_R(C)\;=\;\mathrm{Tr}_R\,U_R(C)\;=\;\chi_R\,\big(U(C)\big),
\end{equation}
where $U_R(C)$ denotes the holonomy evaluated in $R$ and $\chi_R$ is called the corresponding character.\footnote{For a unitary representation, $U_R(C^{-1})=U_R(C)^{-1}=U_R(C)^\dagger$, and therefore
$W_R(C^{-1})=\overline{W_R(C)}$. Equality $W_R(C^{-1})=W_R(C)$ holds when the character is real, e.g.,\ in real/pseudoreal representations such as the defining representation of $SU(2)$.}
In what follows we will keep track of the distinction between the untraced holonomy $U(C)$ and the traced loop $W_R(C)$ that will be essential for reconstruction,
since they encode different amounts of information and play different roles in reconstruction questions: the former retains matrix valued holonomy information, whereas the latter retains only the character of the holonomy in the representation $R$.


\section{Structural properties of Wilson lines and loops}\label{sec:structure}

In Section~\ref{sec:first} we defined the Wilson line $U(\gamma)$ as the unique solution of a first order
transport equation along a path. We now collect the structural properties that follow from this definition and
will be used throughout the reconstruction discussion.

\subsection{Identity, composition, inverses, reparametrization, and path ordering}

Throughout, $\gamma:[0,1]\to M$ is a (piecewise) smooth path and we use the abbreviation
$A(\sigma)\equiv \dot\gamma^\mu(\sigma)\,A_\mu(\gamma(\sigma))$.
When needed we write $U(\gamma;\sigma_2,\sigma_1)$ for the transporter along $\gamma$ from $\sigma_1$ to $\sigma_2$
(as introduced in Section~\ref{sec:first}).

\subsubsection{Identity and composition}

\paragraph{Identity path.}
Let $\mathrm{id}_x$ denote the constant path at $x$. Since no transport occurs along this path,
\begin{equation}\label{eq:identity_path}
U(\mathrm{id}_x)=\mathbbm{1}.
\end{equation}

\paragraph{Segment composition along a fixed path.}

For $0\le \sigma_0\le \sigma_1\le \sigma_2\le 1$, parallel transport along $\gamma$ satisfies the composition law
\begin{equation}\label{eq:segment_comp}
U(\gamma;\sigma_2,\sigma_0)=U(\gamma;\sigma_2,\sigma_1)\,
U(\gamma;\sigma_1,\sigma_0).
\end{equation}
Thus transporting first from $\sigma_0$ to $\sigma_1$ and then from $\sigma_1$ to $\sigma_2$ is equivalent to
transporting directly from $\sigma_0$ to $\sigma_2$ along the same curve.

\paragraph{Path composition (concatenation).}

Let $\gamma_1:[0,1]\to M$ be a path from $y$ to $z$ and $\gamma_2:[0,1]\to M$ a path from $z$ to $x$.
Their concatenation $\gamma_2\circ\gamma_1$ is the path from $y$ to $x$ obtained by following $\gamma_1$ and then
$\gamma_2$ (with any standard piecewise reparametrization).
The parallel transport operator respects concatenation:
\begin{equation}\label{eq:concat_rule}
U(\gamma_2\circ\gamma_1)=U(\gamma_2)\,U(\gamma_1).
\end{equation}
This is the fundamental product rule for parallel transport. A vector at $y$ is first transported to $z$ by $U(\gamma_1)$ and then from $z$ to $x$ by $U(\gamma_2)$; therefore, the total action is $U(\gamma_2)U(\gamma_1)$, with the transporter for the later segment appearing on the left. Although the integral of the connection is additive under the same decomposition,
\begin{equation}
\int_{\gamma_2\circ\gamma_1} A=
\int_{\gamma_1}A+\int_{\gamma_2}A,
\end{equation}
the matrices contributed by the two segments need not commute. Therefore, in a non-Abelian theory, the exponential of the summed integrals cannot in general be treated as an ordinary exponential. Indeed, path ordering is what preserves the sequence of transport and converts the decomposition of the path into the ordered product in Equation \eqref{eq:concat_rule}.
\subsubsection{Inverse paths}

Given a path $\gamma:y\to x$, define the reversed path $\gamma^{-1}:x\to y$ by $\gamma^{-1}(\sigma)=\gamma(1-\sigma)$. Then the corresponding Wilson line is the inverse group element:
\begin{equation}\label{eq:inverse_path}
U(\gamma^{-1}) = U(\gamma)^{-1}.
\end{equation}
Together with the concatenation law, this implies that immediate backtracking contributes no net transport:
\begin{equation}\label{eq:backtrak}
U(\gamma^{-1}\circ\gamma)=U(\gamma)^{-1}U(\gamma)
=\mathbbm{1}.
\end{equation}
This cancellation will be used later when open paths are represented by loops based at a common point.

\subsubsection{Reparametrization invariance}

A Wilson line is invariant under an orientation preserving reparametrization of a fixed path: it does not depend on
the speed with which the path is traversed, although it does retain the orientation and order of traversal.
Concretely, if $f:[0,1]\to[0,1]$ is smooth and strictly increasing, with $f(0)=0$ and $f(1)=1$, then
\begin{equation}\label{eq:reparam_invariance}
U(\gamma\circ f)=U(\gamma).
\end{equation}
This follows from the transport equation, or equivalently from the path ordered exponential, by a change of variables.
Reversal of orientation is not an orientation preserving reparametrization and instead gives the inverse transporter,
as stated in Equation \eqref{eq:inverse_path}.

\subsubsection{Path ordering and the correct factorization identity}

In a non-Abelian gauge theory, the matrices $A(\sigma)$ evaluated at different points along the path generally do
not commute. As shown in \eqref{eq:path_ordered_exp}, the transport equation is therefore represented by a
path-ordered exponential, with larger parameter values placed to the left. This ordering records the sequence in
which the infinitesimal transports are encountered along the path. If the matrices commute for all parameter values,
the ordering may be omitted; this is automatically the case in an abelian theory.

The fundamental factorization property of ordered exponentials is interval splitting (equivalently, segment composition).
If $0\le \sigma_0\le \sigma_1\le \sigma_2\le 1$, then
\begin{equation}\label{eq:interval_split}
\mathcal{P}\exp\left(i\int_{\sigma_0}^{\sigma_2} d\sigma\,A(\sigma)\right)
=\mathcal{P}\exp\left(i\int_{\sigma_1}^{\sigma_2} d\sigma\,A(\sigma)\right)\;
\mathcal{P}\exp\left(i\int_{\sigma_0}^{\sigma_1} d\sigma\,A(\sigma)\right).
\end{equation}
Equation~\eqref{eq:interval_split} is the appropriate factorization law for ordered exponentials. It replaces the
naive identity $e^{A+B}=e^A e^B$, which is generally false when $A$ and $B$ do not commute. When all matrices along
the path commute, path ordering becomes unnecessary and the familiar exponential factorization is recovered
(see, e.g.,\ \cite{Polyakov:1987gfs,Makeenko:2002mcgt}).

\medskip

\paragraph{Example: two constant segments.}
Consider a path composed of two segments, each of parameter length $\Delta$, and suppose that the pulled-back
connection $A(\sigma)$ is constant on each segment, with values $A^{(1)}$ and $A^{(2)}$, respectively.
The concatenation law gives
\begin{equation}
U(\gamma)=\exp\,\big(i\Delta A^{(2)}\big)\,
\exp\,\big(i\Delta A^{(1)}\big),
\end{equation}
where the factor corresponding to the segment traversed later appears on the left. If
$[A^{(1)},A^{(2)}]\neq0$, this product cannot in general be replaced by a single exponential of the sum:
\begin{equation}
\exp\,\big(i\Delta A^{(2)}\big)\,
\exp\,\big(i\Delta A^{(1)}\big)\neq
\exp\,\big(i\Delta(A^{(1)}+A^{(2)})\big).
\end{equation}
Indeed, the Baker-Campbell-Hausdorff expansion makes the first ordering correction explicit:
\begin{equation}
\exp\,\big(i\Delta A^{(2)}\big)\,
\exp\,\big(i\Delta A^{(1)}\big)=\exp\,\left(
i\Delta(A^{(1)}+A^{(2)})+\frac{\Delta^2}{2}[A^{(1)},A^{(2)}]+\mathcal{O}(\Delta^3)\right).
\end{equation}
Thus path ordering retains information about the noncommuting sequence of infinitesimal transports, not merely their sum.




\subsection{First order path deformations and curvature insertion}

Let $\gamma_\epsilon:[0,1]\to M$ be a smooth one-parameter family of paths with fixed endpoints, $\gamma_\epsilon(0)=y$ and $\gamma_\epsilon(1)=x$ for all $\epsilon$, and denote the variation field by
\begin{equation}
u^\mu(\sigma)\equiv \left.\frac{\partial}{\partial\epsilon}\gamma_\epsilon^\mu(\sigma)\right|_{\epsilon=0},
\qquad u^\mu(0)=u^\mu(1)=0.
\end{equation}
We write $\gamma\equiv\gamma_0$ for the undeformed path and abbreviate
\begin{equation}
U(\sigma_2,\sigma_1)\equiv U(\gamma;\sigma_2,\sigma_1)
\end{equation}
for parallel transport along $\gamma$ from $\sigma_1$ to $\sigma_2$. In particular,
$U(1,0)=U(\gamma)$. Differentiating the transport equation with respect to $\epsilon$ and integrating by parts gives the fixed endpoint variation formula, often referred to as a Mandelstam type formula,
\begin{equation}\label{eq:mandelstam_variation}
\delta U(\gamma)
\equiv \left.\frac{\partial}{\partial\epsilon}U(\gamma_\epsilon)\right|_{\epsilon=0}
=i\int_0^1 d\sigma\; U(1,\sigma)\,
F_{\mu\nu}(\gamma(\sigma))\, u^\mu(\sigma)\,
\dot\gamma^\nu(\sigma)\, U(\sigma,0),
\end{equation}
where
\begin{equation}\label{eq:Fnona}
F_{\mu\nu}=\partial_\mu A_\nu-\partial_\nu A_\mu-i[A_\mu,A_\nu].
\end{equation}
The fixed endpoint conditions $u^\mu(0)=u^\mu(1)=0$ remove the boundary contributions.\footnote{If the
endpoints are allowed to vary, the right hand side of \eqref{eq:mandelstam_variation} acquires the additional terms $i\,u^\mu(1)A_\mu(x)U(\gamma) i\,U(\gamma)u^\mu(0)A_\mu(y)$.}
For standard derivations and applications, see, e.g.,\
\cite{Mandelstam:1968,Makeenko:2002mcgt}. The factors $U(1,\sigma)$ and $U(\sigma,0)$ transport the curvature insertion from the point $\gamma(\sigma)$ to the two endpoints. Consequently, the entire right hand side transforms with the same endpoint factors as $U(\gamma)$ itself. Equation~\eqref{eq:mandelstam_variation} makes two points precise:
\begin{itemize}
\item Shape dependence and curvature.
In general, Wilson lines are not topological invariants: their dependence on the shape of the path is already visible
at first order and is controlled locally by $F_{\mu\nu}$. Equation~\eqref{eq:mandelstam_variation} shows that the response of a Wilson line to a fixed endpoint deformation is
governed locally by the curvature encountered along the path.
In particular, for a flat connection ($F_{\mu\nu}=0$) on a simply connected patch, the transporter is unchanged under
smooth fixed endpoint deformations within that region and consequently depends only on its endpoints. On a
not simply connected region, a flat connection may still carry nontrivial holonomy that depends on the homotopy class
of the path.

\item Tangential deformations and reparametrization.
A purely tangential deformation has the form
$u^\mu(\sigma)=f(\sigma)\dot\gamma^\mu(\sigma)$, with $f(0)=f(1)=0$. Its contribution to
\eqref{eq:mandelstam_variation} vanishes because
\begin{equation}
F_{\mu\nu}\,u^\mu\dot\gamma^\nu=
f(\sigma)F_{\mu\nu}\dot\gamma^\mu\dot\gamma^\nu=0.
\end{equation}
Thus tangential deformations, which change the parametrization without changing the path itself, leave
$U(\gamma)$ unchanged, in agreement with \eqref{eq:reparam_invariance}.
\end{itemize}
For comparison, in an abelian theory path ordering is unnecessary, and the fixed endpoint variation reduces to
\begin{equation}
\delta\,\left(\int_\gamma A_\mu\,dx^\mu\right)
= \int_0^1 d\sigma\; F_{\mu\nu}(\gamma(\sigma))\,
u^\mu(\sigma)\, \dot\gamma^\nu(\sigma).
\end{equation}
The non-Abelian expression \eqref{eq:mandelstam_variation} has the same local curvature refinement supplemented by
parallel transport on both sides so that the result remains gauge covariant at the endpoints. It provides the
infinitesimal basis for small loop curvature expansions and is closely related to non-Abelian Stokes type
representations of loop holonomies
(see, e.g.,\ \cite{DiakonovPetrov:2000nast,HirayamaUeno:2000nast}).

\subsection{Wilson line for a straight segment}\label{subsec:straight_segment}

\begin{figure}[t]
  \centering
  \includegraphics[width=0.6\textwidth]{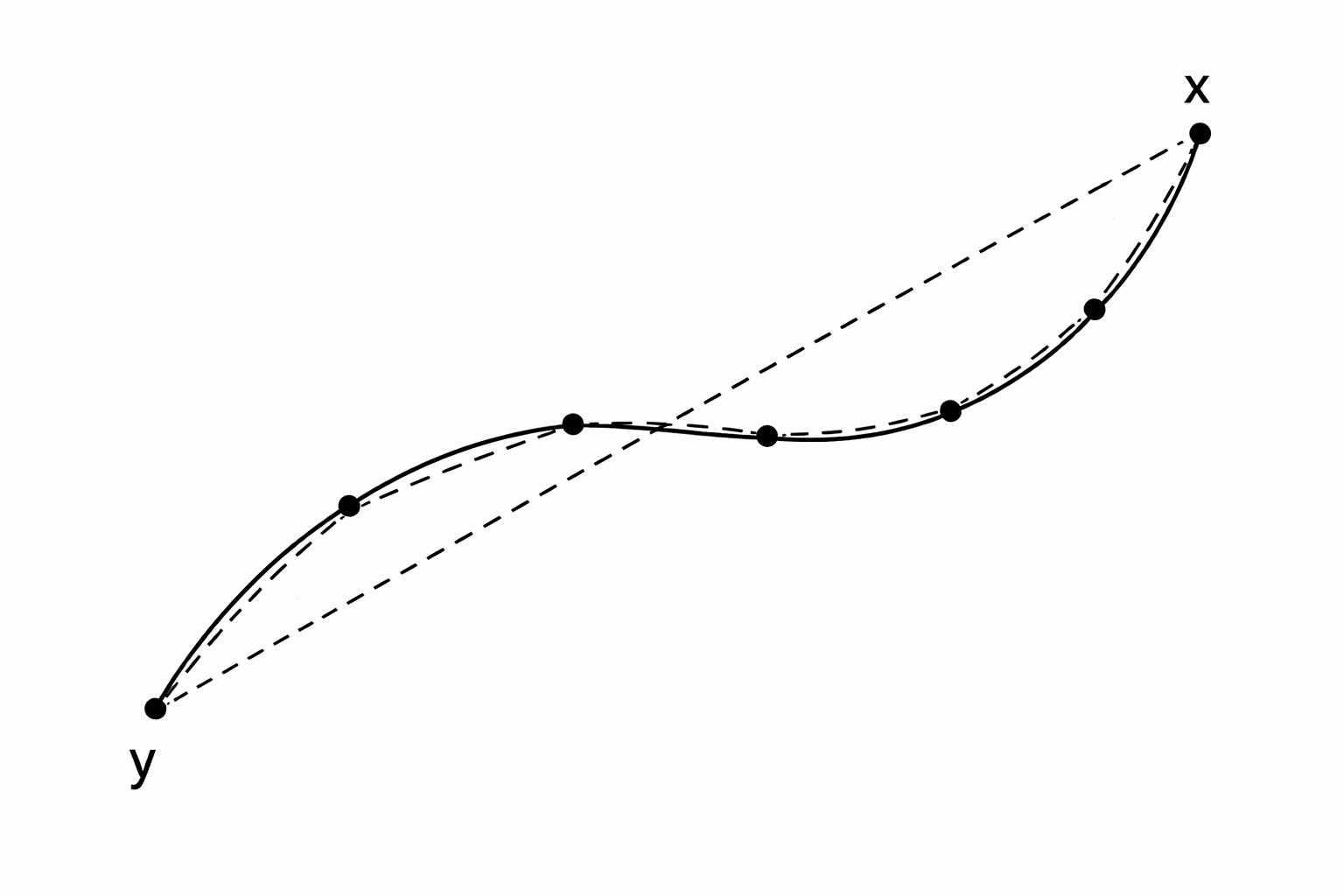}
  \caption{Schematic paths from $y$ to $x$. The solid curve represents a path $\gamma$, the nearby dashed curve a
fixed endpoint deformation $\gamma_\epsilon$, and the straight dashed chord the segment $S_{x,y}$ used for the
small separation expansion. The marked points illustrate a partition of the path used to approximate parallel
transport by an ordered product of short segment transporters.}\label{fig:paths_straight_segment}
\end{figure}

For local expansions, let $x$ and $y$ be sufficiently close points in a coordinate patch and choose the straight
segment from $y$ to $x$,
\begin{equation}
S_{x,y}^{\mu}(\tau)=y^\mu+\tau\,(x^\mu-y^\mu),\qquad \tau\in[0,1],
\end{equation}
so that $S_{x,y}(0)=y$ and $S_{x,y}(1)=x$.
On a general manifold one may instead choose the short geodesic joining the two points; the linear term in the local expansion below is unchanged in a chosen local trivialization. Figure~\ref{fig:paths_straight_segment} summarizes generic the geometric constructions used in the path deformation and
local expansion discussions.

Let $\ell^\mu\equiv x^\mu-y^\mu$. Along the straight segment, $\dot S_{x,y}^{\mu}(\tau)=\ell^\mu$, and the corresponding Wilson line is

\begin{equation}\label{eq:U_straight_def}
U(S_{x,y})=\mathcal{P}\exp\left(i\int_0^1 d\tau\;\ell^\mu\,A_\mu(y+\tau\ell)\right).
\end{equation}
Retaining the first nontrivial term in the Dyson series gives

\begin{equation}\label{eq:U_straight_firstorder_step}
U(S_{x,y})=\mathbbm{1} +i\int_0^1 d\tau\;\ell^\mu\,A_\mu(y+\tau\ell)  +\mathcal{O}(\ell^2),
\end{equation}
where $\mathcal{O}(\ell^2)$ denotes terms of second or higher order in the coordinate displacement $\ell^\mu$.
For a smooth connection,
\begin{equation}
A_\mu(y+\tau\ell)=A_\mu(y)+\mathcal{O}(\ell),
\end{equation}
and therefore
\begin{equation}\label{eq:U_straight_firstorder}
U(S_{x,y}) =\mathbbm{1}+ i\,\ell^\mu A_\mu(y)+\mathcal{O}(\ell^2).
\end{equation}
At linear order only the first Dyson term contributes, so path ordering does not play a role. Beyond linear order, both the variation of $A_\mu$ along the segment and the higher terms of the Dyson series contribute.
Equation~\eqref{eq:U_straight_firstorder} is the basic small separation expansion used in the next subsection to
identify the local gauge potential as the infinitesimal generator of parallel transport in the chosen local
trivialization.


\subsection{Infinitesimal segment and extraction of the connection}\label{subsec:infinitesimal_extraction}

Let $x\in M$ and let $\epsilon^\mu$ be a small displacement in a chosen coordinate patch. Consider the straight segment $S_{x+\epsilon,x}$ from $x$ to $x+\epsilon$. Setting $y=x$ and $\ell^\mu=\epsilon^\mu$ in expression \eqref{eq:U_straight_firstorder} gives
\begin{equation}\label{eq:U_inf_segment}
U(S_{x+\epsilon,x})=\mathbbm{1}+i\,\epsilon^\mu A_\mu(x)+\mathcal{O}(\epsilon^2).
\end{equation}
All quantities in this equation are evaluated in the same chosen representation.

To recover the connection, choose a tangent vector $v^\mu$ at $x$ and consider the one-parameter family of segments
$S_{x+sv,x}$. Substituting $\epsilon^\mu=s\,v^\mu$ into \eqref{eq:U_inf_segment} gives
\begin{equation}
U(S_{x+sv,x})
=\mathbbm{1}+i\,s\,v^\mu A_\mu(x)+\mathcal{O}(s^2).
\end{equation}
It follows that
\begin{equation}\label{eq:Av_directional}
A(v)(x)\equiv\frac{1}{i}\left.
\frac{d}{ds}U(S_{x+sv,x})\right|_{s=0}=v^\mu A_\mu(x).
\end{equation}
Choosing $v=e_\mu$, where $e_\mu$ is a coordinate basis vector, gives
\begin{equation}\label{eq:A_component_from_U}
A_\mu(x)=\frac{1}{i}
\left. \frac{d}{ds}U(S_{x+s e_\mu,x}) \right|_{s=0}.
\end{equation}
Thus, the connection is the infinitesimal generator of parallel transport in the chosen local trivialization.

Note that: 
\begin{enumerate}[label=\alph*)]

\item Gauge transformation. Applying \eqref{eq:gauge_trans_U} to $S_{x+sv,x}$ gives
\begin{equation}
U(S_{x+sv,x})
\longmapsto
g(x+sv)\,U(S_{x+sv,x})\,g(x)^{-1}.
\end{equation}
Differentiating at $s=0$ yields
\begin{align}\label{eq:Av_gauge_trans}
A^{\,g}(v)(x)&=g(x)\,A(v)(x)\,g(x)^{-1}-i\,v^\mu\big(\partial_\mu g(x)\big)g(x)^{-1}\nonumber\\
&=g(x)\,A(v)(x)\,g(x)^{-1}+i\,v^\mu g(x)\,\partial_\mu g(x)^{-1}.
\end{align}
In components,
\begin{align}
A_\mu^{\,g}(x)&=g(x)\,A_\mu(x)\,g(x)^{-1}
-i\,\big(\partial_\mu g(x)\big)g(x)^{-1}\nonumber\\
&=g(x)\,A_\mu(x)\,g(x)^{-1}+i\,g(x)\,\partial_\mu g(x)^{-1},
\end{align}
in agreement with Equation \eqref{eq:gauge_trans_A}.

\item Dependence on the short path. At linear order, the result depends only on the endpoint displacement $\epsilon^\mu$.
Any smooth short curve beginning at $x$ and ending at
$x+s v+\mathcal{O}(s^2)$ gives
\begin{equation}
U=\mathbbm{1}+i\,s\,v^\mu A_\mu(x)+\mathcal{O}(s^2).
\end{equation}
The detailed shape of the curve enters only at higher order. For deformations with fixed endpoints, this shape
dependence is controlled by the curvature insertion formula \eqref{eq:mandelstam_variation}.
\end{enumerate}
The next section develops the converse construction: finite Wilson lines are obtained by composing infinitesimal
transports along a path.
\section{Constructing $U(\gamma)$ from infinitesimal segments}\label{sec:constr}

In Section~\ref{subsec:infinitesimal_extraction} we 
showed how the connection is recovered from an infinitesimal Wilson line. We now reverse the construction: starting from $A_\mu(x)$ in a chosen local trivialization, we build parallel transport along a finite path by composing short segment transporters. Let $\gamma:[0,1]\to M$ be a piecewise smooth path with
\begin{equation}
\gamma(0)=y, \qquad \gamma(1)=x.
\end{equation}
And choose a partition $\pi: 0=\tau_0<\tau_1<\cdots<\tau_N=1,$ and set
\begin{equation}
\Delta\tau_i:=\tau_{i+1}-\tau_i,
\qquad \gamma_i:=\gamma|_{[\tau_i,\tau_{i+1}]}.
\end{equation}
For a sufficiently fine partition, the infinitesimal expansion on each segment gives
\begin{equation}\label{eq:segment-expansion}
U(\gamma_i)
= \mathbbm{1} +i\,\Delta\tau_i\,
\dot\gamma^\mu(\tau_i)\,
A_\mu(\gamma(\tau_i))
+\mathcal{O}(\Delta\tau_i^2).
\end{equation}
For a smooth connection and path, the local errors vanish in the continuum limit: indeed,
\begin{equation}
\sum_i(\Delta\tau_i)^2\leq |\pi|\sum_i\Delta\tau_i=|\pi|, \qquad |\pi|:=\max_i \Delta\tau_i.    
\end{equation}
Using the concatenation (product) property of parallel transport, the Wilson line along the full path is obtained as the ordered product of segment transports,
\begin{equation}\label{eq:product-limit}
U(\gamma)=\lim_{|\pi|\to 0}\;
U(\gamma_{N-1})\,U(\gamma_{N-2})\cdots U(\gamma_0).
\end{equation}
A field at $y$ is acted on first by $U(\gamma_0)$, then by $U(\gamma_1)$, and so on. Therefore, the factor associated
with the segment traversed later appears farther to the left in \eqref{eq:product-limit}. In a non-Abelian theory
these factors generally do not commute, so their order must be retained as the partition is refined. This ordered
continuum limit is the origin of the path ordering symbol $\mathcal{P}$.

The ordered product limit satisfies the transport equation
\eqref{eq:parallel_transport_ode} with the initial condition
$U(\gamma;0,0)=\mathbbm{1}$. By uniqueness of the solution, it is the same Wilson line defined in Section~\ref{sec:first}. It may therefore be written as
\begin{equation}\label{eq:path_ordered_exp_construction}
U(\gamma)=\mathcal{P}\exp\left(i\int_0^1 d\tau\;
\dot\gamma^\mu(\tau)\,A_\mu(\gamma(\tau))\right)
=\mathcal{P}\exp\left(i\int_\gamma A\right),
\end{equation}
where $A=A_\mu\,dx^\mu$.

Thus the path ordered exponential is the continuum limit of an ordered product of infinitesimal parallel
transporters.


\section{Base point transport and covariant derivatives of loop holonomies}\label{sec:covariant}

Let $C_x$ denote a smooth family of closed loops based at $x$; for example, within a coordinate patch one may
translate a fixed loop shape together with its base point. Let $U(C_x)$ be the corresponding untraced holonomy in a
chosen representation. Since $U(C_x)$ and $U(C_{x+\epsilon})$ are based at different points and transform with different endpoint factors, they must first be transported to a common base point before they can be compared.

Let $\epsilon^\mu$ be a small displacement and let $S_{x+\epsilon,x}$ be the straight segment from $x$ to $x+\epsilon$,
with inverse segment $S_{x,x+\epsilon}=(S_{x+\epsilon,x})^{-1}$.
Define the displaced holonomy transported back to the base point $x$ by
\begin{equation}\label{eq:transported_displaced_loop}
\widetilde U(C_{x+\epsilon};x)
\;:=\;
U(S_{x,x+\epsilon})\,U(C_{x+\epsilon})\,U(S_{x+\epsilon,x}).
\end{equation}
By construction, $\widetilde U(C_{x+\epsilon};x)$ transforms covariantly at the single base point $x$,
\begin{equation}
\widetilde U(C_{x+\epsilon};x)\ \longmapsto\ g(x)\,\widetilde U(C_{x+\epsilon};x)\,g(x)^{-1}.
\end{equation}
Using the infinitesimal segment expansion \eqref{eq:U_inf_segment} and the inverse path relation
\eqref{eq:inverse_path}, we have
\begin{equation}
\begin{aligned}
U(S_{x+\epsilon,x})
&=\mathbbm{1}+i\,\epsilon^\mu A_\mu(x)
+\mathcal{O}(\epsilon^2),\\
U(S_{x,x+\epsilon})&=
U(S_{x+\epsilon,x})^{-1}=\mathbbm{1}
-i\,\epsilon^\mu A_\mu(x)+\mathcal{O}(\epsilon^2).
\end{aligned}
\end{equation}
Substituting into Equation \eqref{eq:transported_displaced_loop} and expanding to first order gives
\begin{equation}\label{eq:first_step_displaced}
\widetilde U(C_{x+\epsilon};x)
=U(C_{x+\epsilon})
- i\,\epsilon^\mu\,[A_\mu(x),U(C_{x+\epsilon})]+\mathcal{O}(\epsilon^2).
\end{equation}
Assuming that both the connection and the chosen family $C_x$ vary smoothly with $x$, we may expand
\begin{equation}
U(C_{x+\epsilon}) = U(C_x) + \epsilon^\mu \partial_\mu U(C_x) + \mathcal{O}(\epsilon^2).
\end{equation}
Here $\partial_\mu U(C_x)$ denotes the change in the holonomy produced by varying the loop within the chosen family $C_x$; it is not a derivative along the loop parameter. Substituting into Equation \eqref{eq:first_step_displaced} yields the gauge covariant Taylor expansion
\begin{equation}\label{eq:covariant_derivative_holonomy}
\widetilde U(C_{x+\epsilon};x)
=U(C_x)+\epsilon^\mu\Big(\partial_\mu U(C_x)- i[A_\mu(x),U(C_x)]\Big)+\mathcal{O}(\epsilon^2).
\end{equation}
It is natural to recognize the combination in parentheses as the adjoint covariant derivative at the base point,
\begin{equation}\label{eq:adjoint_covariant_derivative}
D_\mu U(C_x)\;:=\;\partial_\mu U(C_x)- i[A_\mu(x),U(C_x)].
\end{equation}
Thus,
\begin{equation}
\widetilde U(C_{x+\epsilon};x)
=U(C_x)+\epsilon^\mu D_\mu U(C_x)+\mathcal{O}(\epsilon^2).
\end{equation}
The commutator term in relation \eqref{eq:adjoint_covariant_derivative} compensates for the position dependence of the conjugating gauge transformation, so
\begin{equation}
D_\mu U(C_x) \longmapsto g(x)\,D_\mu U(C_x)\,g(x)^{-1}.
\end{equation}

Note that:
\begin{enumerate}[label=\alph*)]
\item Abelian case. In an abelian theory the commutator term vanishes and therefore $D_\mu U(C_x)=\partial_\mu U(C_x)$.

\item Relation to curvature. The two short segments in Equation \eqref{eq:transported_displaced_loop} produce the commutator term with $A_\mu(x)$. The remaining derivative $\partial_\mu U(C_x)$ describes how the holonomy changes as the chosen loop is moved or deformed. That change can in turn be expressed through the curvature insertion formula
\eqref{eq:mandelstam_variation}. Thus, the present construction and the path-deformation formula describe two
parts of the same comparison: first transport the displaced holonomy to a common base point, and then evaluate
how the loop itself has changed.

\item Taking the trace. Taking a trace removes the conjugation in Equation \eqref{eq:transported_displaced_loop}, so
$\mathrm{Tr}_R\,\widetilde U_R(C_{x+\epsilon};x) =\mathrm{Tr}_R\,U_R(C_{x+\epsilon}).$.
Equivalently, $\partial_\mu W_R(C_x)=\mathrm{Tr}_R\,\big(D_\mu U_R(C_x)\big)$ since the trace eliminates commutators.
\end{enumerate}

\subsection{Changing the base point of a loop}\label{subsec:basepoint_change}


A loop holonomy is attached to a chosen starting point. For a loop $C_y:y\to y$, the operator $U(C_y)$ transports
a field once around the loop and back to $y$, and under a gauge transformation it obeys
\begin{equation}
U(C_y)\longmapsto g(y)\,U(C_y)\,g(y)^{-1}.
\end{equation}
A loop based at $x$ transforms instead with $g(x)$. Holonomies based at different points therefore cannot be
compared directly; one must choose a path connecting the two base points and use its Wilson line to transport one
holonomy to the other point.

Let $Q_{y,x}$ be any path from $y$ to $x$ (so $Q_{y,x}:y\to x$) and let $C_y$ be a loop based at $y$. The corresponding loop based at $x$ obtained by transporting $C_y$ along $Q_{y,x}$ is
\begin{equation}\label{eq:rebase_loop}
C_x \;\equiv\; Q_{y,x}\circ C_y \circ Q_{y,x}^{-1}.
\end{equation}
Using concatenation and inversion properties of parallel transport, its holonomy is
\begin{equation}\label{eq:rebase_holonomy}
U(C_x)=U(Q_{y,x})\,U(C_y)\,U(Q_{y,x})^{-1}.
\end{equation}
Equation~\eqref{eq:rebase_holonomy} shows  changing the base point of a loop changes its holonomy by conjugation.
If $Q_{y,x}$ is chosen along the original loop, this simply changes the point at which the traversal begins.
For a general connecting path, the construction attaches a path and its inverse; the resulting
holonomy is still conjugate to $U(C_y)$.

The relation is consistent with gauge covariance because the endpoint factors in
$U(Q_{y,x})$ cancel those associated with $U(C_y)$. Taking a trace removes the remaining conjugation, so the traced Wilson loop is base point independent:
\begin{equation}\label{eq:trace_basepoint_independence}
W_R(C_x)=\mathrm{Tr}_R\,U_R(C_x)
=\mathrm{Tr}_R\,U_R(C_y)=W_R(C_y).
\end{equation}
Thus the untraced holonomy changes by conjugation when the base point is changed, whereas its character does not.
In particular, the traced Wilson loop associated with a closed contour is independent of the point chosen to begin
the traversal.

\medskip
\noindent

\paragraph*{Example:}
 Small rectangular loop. Having established that the traced Wilson loop is independent of the point chosen to begin the traversal, we may choose any convenient base point on a closed contour. Let $x$ be one corner of a small square
$C_{\mu\nu}(x;a)$ of side length $a$ in the $\mu\nu$-plane, with the orientation chosen so that its leading area
element is $+a^2$ in that plane. For a smooth connection, the corresponding holonomy has the expansion
\begin{equation}\label{eq:small_loop_expansion}
U(C_{\mu\nu}(x;a))=\mathbbm{1}+ i\,a^2\,F_{\mu\nu}(x)+\mathcal{O}(a^3), \qquad
F_{\mu\nu}=\partial_\mu A_\nu-\partial_\nu A_\mu - i[A_\mu,A_\nu],
\end{equation}
so the leading deviation of the holonomy from the identity encodes the curvature. For a semisimple gauge group, $\mathrm{Tr}_R(F_{\mu\nu})=0$ in every finite dimension representation, so the term linear in the curvature disappears after taking the trace. The first nonzero correction is therefore\footnote{If the gauge group contains an abelian factor, $\mathrm{Tr}_R(F_{\mu\nu})$ need not vanish, and the traced loop may already
contain a correction of order $a^2$.}
\begin{equation}
\mathrm{Tr}_R\,U_R(C_{\mu\nu}(x;a))=\dim(R) -\frac{a^4}{2}\,\mathrm{Tr}_R\,\big(F_{\mu\nu}(x)^2\big) +\cdots.
\end{equation}
If the same rectangular contour is instead based at another point $y$ on its boundary, the untraced holonomy is
conjugated according to relation \eqref{eq:rebase_holonomy}, while its trace and the expansion above remain unchanged, in accordance with expression \eqref{eq:trace_basepoint_independence}.


\section{Recovering open holonomies from based loop data}\label{sec:recover}

In this section we work at the holonomy level. We assume that matrix valued holonomies $U(C)$ are known for loops
$C$ based at a fixed point $x$, in a chosen representation. To recover transport between arbitrary points, this
based loop data must be supplemented by a choice of reference transports from $x$ to the other points in the region.

This construction cannot be done with a single traced Wilson loop functional alone: in general,
$W_R(C)=\mathrm{Tr}_R\,U_R(C)$ does not determine the matrix $U_R(C)$. We return to this distinction in
Section~\ref{sec:reconst}; see also \cite{Giles:1981reconstruct,Barrett:1991holonomy}.

\subsection{Reference paths and the reconstruction formula}

Fix a base point $x$ in a region of interest and, for every point $y$ in that region, choose a reference path
$\pi_{x,y}:x\to y$, with $\pi_{x,x}=\mathrm{id}_x$. Associate with it a reference transporter
\begin{equation}\label{eq:reference_holonomy}
G(y):x\to y, \qquad G(x)=\mathbbm{1}.
\end{equation}
If an underlying connection is already given, then $G(y)=U(\pi_{x,y})$. Note that in the reconstruction problem, however, the reference transports are additional choices and are not determined by the based loop holonomies alone.

Now let $P_{y,z}:y\to z$ be an arbitrary open path. Together with the two reference paths it defines a loop based at
$x$:
\begin{equation}\label{eq:based_loop_from_open_path}
C_x(P_{y,z}) \;\equiv\; \pi_{x,z}^{-1}\circ P_{y,z}\circ \pi_{x,y}.
\end{equation}
This loop starts at $x$, goes to $y$ along $\pi_{x,y}$, then follows $P_{y,z}$ to $z$, and finally returns to $x$
along $\pi_{x,z}^{-1}$. Using the concatenation and inverse path properties of parallel transport,
\begin{equation}\label{eq:loop_from_open_path_holonomy}
U\,\left(C_x(P_{y,z})\right)
=U(\pi_{x,z}^{-1})\,U(P_{y,z})\,U(\pi_{x,y})
=G(z)^{-1}\,U(P_{y,z})\,G(y).
\end{equation}
Solving this relation for the open path holonomy gives the reconstruction formula
\begin{equation}\label{eq:open_from_loop}
U(P_{y,z})
=G(z)\,U\,\left(C_x(P_{y,z})\right)\,G(y)^{-1}.
\end{equation}
Equation~\eqref{eq:open_from_loop} shows that based loop holonomies at $x$, together with the chosen reference
transports $G(y)$, determine the transporter along every open path $P_{y,z}$. In fact, the role of the reference paths is to turn the open path into a loop based at $x$, while the factors $G(y)$ and $G(z)$ restore the original endpoints.


\subsection{Reference path ambiguity and its gauge-like action}
\label{subsec:refpath_ambiguity}

The reconstruction formula \eqref{eq:open_from_loop} depends on a choice of reference paths
$y\mapsto\pi_{x,y}$ and their holonomies $G(y)=U(\pi_{x,y})$.
This choice is not unique. One may choose a different family $\pi'_{x,y}$ from $x$ to $y$, with
\begin{equation}
G'(y)=U(\pi'_{x,y}).
\end{equation}
The two choices are related by the point dependent group element
\begin{equation}\label{eq:h_def}
H(y)
\equiv
U\,\left(\pi'_{x,y}\circ\pi_{x,y}^{-1}\right)=G'(y)\,G(y)^{-1},
\end{equation}
where $\pi'_{x,y}\circ\pi_{x,y}^{-1}$ is a comparison loop based at $y$. Equivalently,
\begin{equation}\label{eq:G_change}
G'(y)=H(y)\,G(y).
\end{equation}
With the natural convention
$\pi_{x,x}=\pi'_{x,x}=\mathrm{id}_x$, one has $H(x)=\mathbbm{1}$.

Changing the reference paths also changes the based loop associated with an open path:
\begin{equation}
C_x^{\,\pi}(P_{y,z})
=\pi_{x,z}^{-1}\circ P_{y,z}\circ\pi_{x,y}, \qquad
C_x^{\,\pi'}(P_{y,z}) =\pi_{x,z}'{}^{-1}\circ P_{y,z}\circ\pi'_{x,y}.
\end{equation}
If all holonomies arise from the same connection, these two changes cancel each other:
\begin{equation}\label{eq:reference_path_consistency}
G'(z)\, U\,\left(C_x^{\,\pi'}(P_{y,z})\right)\,
G'(y)^{-1} = G(z)\,
U\,\left(C_x^{\,\pi}(P_{y,z})\right)\,
G(y)^{-1} = U(P_{y,z}).
\end{equation}
Thus changing the reference paths does not change the open holonomy when the corresponding based loop and reference
holonomies are changed consistently.


In the reconstruction problem, however, the based loop holonomies at $x$ do not by themselves fix the reference
holonomies $G(y)$. If another choice is made,
\begin{equation}
\widetilde G(y)=h(y)\,G(y), \qquad h(x)=\mathbbm{1},
\end{equation}
while the based loop data at $x$ are kept fixed, then the reconstructed open holonomy changes as
\begin{equation}\label{eq:U_change_reference}
U_{\widetilde G}(P_{y,z}) = h(z)\,U_G(P_{y,z})\,h(y)^{-1}.
\end{equation}
This has exactly the same endpoint structure as the gauge transformation law for an open Wilson line.
For a closed loop, it acts by conjugation at the loop's base point, so characters and traces are unchanged.
For loops based at the fixed point $x$, even the untraced holonomy is unchanged because $h(x)=\mathbbm{1}$.
Therefore the freedom in choosing the reference holonomies is gauge like: different choices give gauge equivalent
open holonomies while leaving the given based loop data unchanged.

\subsection{Consistency with concatenation}

The reconstructed open transport must satisfy the same composition law as ordinary parallel transport.
Let $P_{y,z}:y\to z$ and $P_{z,w}:z\to w$ be composable paths. Using Equation \eqref{eq:open_from_loop},

\begin{align}
U(P_{z,w})\,U(P_{y,z})
&=\Bigl(G(w)\,U(C_x(P_{z,w}))\,G(z)^{-1}\Bigr)
\Bigl(G(z)\,U(C_x(P_{y,z}))\,G(y)^{-1}\Bigr)\nonumber\\
&=G(w)\,U(C_x(P_{z,w}))\,U(C_x(P_{y,z}))\,G(y)^{-1}.
\end{align}
The product of the two based loops contains an immediate retracing of the reference path $\pi_{x,z}$:
\begin{equation}
C_x(P_{z,w})\circ C_x(P_{y,z})
=\pi_{x,w}^{-1}\circ P_{z,w}\circ
\underbrace{\pi_{x,z}\circ\pi_{x,z}^{-1}}_{\text{immediate backtracking}} \circ P_{y,z}\circ\pi_{x,y}.
\end{equation}
By the backtracking relation \eqref{eq:backtrak}, this retraced part contributes the identity.
Therefore,

\begin{equation}
U\,\left(C_x(P_{z,w})\right)\,U\,\left(C_x(P_{y,z})\right)
=U\,\left(C_x(P_{z,w}\circ P_{y,z})\right).
\end{equation}
Substituting this relation into the reconstructed product gives
\begin{equation}
U(P_{z,w}\circ P_{y,z}) \;=\; U(P_{z,w})\,U(P_{y,z}).
\end{equation}
Therefore, the reconstruction formula preserves path composition, as required.
\section{Reconstructing gauge fields from loop data}\label{sec:reconst}

\subsection{Precise statement of the reconstruction problem}
Let $M$ be a smooth manifold and $G$ a Lie group. As in
Section~\ref{subsec:global_patching}, we work locally on a single trivializing patch -in a region where the principal bundle may be
taken as trivial- and write the connection as the
$\mathfrak{g}$-valued one form $A=A_\mu(x)\,dx^\mu$. We adopt the same endpoint convention as for the straight segments $S_{x,y}$ used earlier: a path $\gamma_{x,y}$ is a
piecewise smooth oriented map $\gamma:[0,1]\to M$ with
\begin{equation}
\gamma(0)=y,\qquad \gamma(1)=x.
\end{equation}
Under a gauge transformation $g:M\to G$, the open path holonomy transforms at its endpoints as
\begin{equation}
U_A(\gamma_{x,y}) \longmapsto g(x)\,U_A(\gamma_{x,y})\,g(y)^{-1}.
\end{equation}
And for a based loop $C_x:x\to x$, this reduces to conjugation:
\begin{equation}\label{eq:U_conjugate}
U_A(C_x) \longmapsto g(x)\,U_A(C_x)\,g(x)^{-1}.
\end{equation}
For an individual loop, a gauge transformation changes $U_A(C_x)$ only by conjugation. Therefore the gauge invariant information carried by this matrix is its conjugacy class; equivalently, it is the information that
does not change under \eqref{eq:U_conjugate}. For several loops $C_1,C_2,\ldots$ based at the same point $x$, the same group element $g(x)$ conjugates all of
their holonomies simultaneously. This means that one may also form joint gauge invariant quantities, such as
\begin{equation}
\mathrm{Tr}_R\,\big(U_{A,R}(C_1)U_{A,R}(C_2)\big),
\end{equation}
which contain information about how the holonomies are related to one another and cannot, in general, be recovered
from the separate traces $\mathrm{Tr}_R U_{A,R}(C_1)$ and $\mathrm{Tr}_R U_{A,R}(C_2)$ alone. These properties form the basis of the holonomy reconstruction problem discussed in
\cite{Giles:1981reconstruct,Barrett:1991holonomy}; see also \cite{KobayashiNomizu:1963fdg1}.

The traced Wilson loop in a representation $R$ is the corresponding character, $W_R(C)\;:=\;\mathrm{Tr}_R\,U_{A,R}(C),$ which is invariant under conjugation and therefore independent of the point chosen to begin the traversal of $C$.
\medskip
\noindent

\medskip
\noindent
\textbf{Reconstruction question.}
Given loop space data $\mathcal{W}$ - for example, the values $\{W_R(C)\}$ for all closed loops $C$ and a specified
set of representations $R$ - does there exist a connection $A$ that generates these data? If so, to what extent is
$A$ uniquely determined up to gauge equivalence?
\medskip
\noindent

 There are two related inverse questions:
 
\medskip
\noindent

(i) \emph{Holonomy level data.}
At the holonomy level, one is given a candidate assignment
$\gamma\mapsto U(\gamma)$ and asks whether it arises from a smooth connection $A$, and whether that connection is
unique up to gauge. More preciesly, a group element $U(\gamma)\in G$ is assigned to each piecewise smooth path $\gamma$, encoding the corresponding parallel transport. Because this assignment retains the untraced transport information, it is the natural input for
direct reconstruction of the connection.

(ii) \emph{Traced Wilson loop data.} At the traced level, one is instead given character data\footnote{Referring to the collection of traced Wilson loops in one or more representations.}
\begin{equation}
C\longmapsto W_R(C)
\end{equation}
for a specified set of representations and asks whether it can first be lead to a compatible holonomy assignment
and then to a connection. Therefore, 
A complex number $W_R(C)=\mathrm{Tr}_R\,U_R(C)$ is assigned to each closed loop $C$ and representation $R$. These quantities are gauge invariant, but this second case is more restrictive because taking the trace removes part of the holonomy information.

Holonomy data is therefore the natural input for direct reconstruction of the connection, whereas traced Wilson loop data must satisfy additional algebraic, positivity, and regularity conditions before such a reconstruction can be attempted.\footnote{Throughout,
reconstruction refers to the classical smooth problem on a local patch: recovering $A_\mu(x)$ up to gauge from
compatible holonomy data under suitable regularity assumptions. Generalized or distributional connections used in some loop and quantum representations obey different continuity and admissibility conditions and are not considered here.}
The next subsection addresses consistency at the holonomy level. The subsequent subsections turn to the additional questions that arise when only traced loop data is available.
\subsection{Holonomy level consistency}

Not every assignment of group elements to paths can arise as the holonomy of a connection. Indeed, holonomy data must satisfy basic compatibility conditions under operations on paths, together with suitable regularity conditions.
Concretely, a connection $A$ determines a map
$\gamma\mapsto U_A(\gamma)\in G$ for each piecewise smooth oriented path $\gamma_{x,y}$, with $\gamma(0)=y$ and $\gamma(1)=x$; in any chosen representation, its matrix is given by the path ordered exponential (see, e.g.,\ \cite{Nakahara:2003,BaezMuniain:1994}). Abstractly, the assignment $\gamma\mapsto U(\gamma)$ must satisfy the following consistency conditions
(see \cite{Barrett:1991holonomy} for an axiomatic formulation and reconstruction viewpoint):

\begin{enumerate}[label=(\roman*)]
\item Identity path. The constant path at $x$ carries no transport:
\begin{equation}
U(\mathrm{id}_x)=\mathbbm{1}.
\end{equation}

\item Concatenation. Stepwise transport composes: if $\eta_{y,z}$ runs from $z$ to $y$ and $\gamma_{x,y}$
runs from $y$ to $x$, then
\begin{equation}
U(\gamma_{x,y}\circ \eta_{y,z})=U(\gamma_{x,y})\,U(\eta_{y,z}).
\end{equation}

\item Inverse paths. Reversing orientation inverts transport:
\begin{equation}
U(\gamma^{-1})=U(\gamma)^{-1}.
\end{equation}

\item Reparametrization invariance. The holonomy is unchanged under an orientation preserving reparametrization of the path, i.e., $U(\gamma)$ depends on the oriented curve, not on the particular monotone
parametrization used to describe it.

\item Local regularity.
The holonomy must vary smoothly when the path is varied smoothly. In particular, for a short coordinate segment,
the map
\begin{equation}
s\longmapsto U(S_{x+s e_\mu,x})
\end{equation}
must be differentiable at $s=0$. Its derivative defines the local connection, by \eqref{eq:A_component_from_U}, and therefore gives the short segment expansion
\begin{equation}
U(S_{x+\epsilon,x})
= \mathbbm{1} +i\,\epsilon^\mu A_\mu(x)
+\mathcal{O}(\epsilon^2),
\end{equation}
as discussed in Section~\ref{subsec:infinitesimal_extraction}, so that $A_\mu(x)$ is recovered from the $\epsilon\to 0$ limit.


\end{enumerate}
Finally, one should remember that the connection reconstructed from holonomy data is determined only up to gauge: if $g:M\to G$, then the transformed assignment
\begin{equation}
U(\gamma_{x,y})\ \mapsto\ g(x)\,U(\gamma_{x,y})\,g(y)^{-1}
\end{equation}
produces holonomy data describing a gauge equivalent connection. With the compatibility and regularity conditions above stated precisely, a consistent holonomy assignment determines a local connection up to gauge; see
\cite{Giles:1981reconstruct,Barrett:1991holonomy}.

\subsection{Wilson loop functionals}

For a based loop $C_{x_0}:x_0\to x_0$, the holonomy transforms by conjugation under a gauge transformation,
$U(C_{x_0})\mapsto g(x_0)\,U(C_{x_0})\,g(x_0)^{-1}$.
Moreover, changing the base point along the same geometric loop also conjugates the holonomy (cf.\ the discussion around Equation
\eqref{eq:rebase_holonomy}). Since the trace is invariant under conjugation, the traced Wilson loop , $W_R(C)\;=\;\mathrm{Tr}_R\,U_R(C)$, is gauge invariant and independent of the chosen base point on $C$\footnote{Here and throughout, $W_R(C)$ denotes the unnormalized trace. Some conventions instead include a factor
$1/\dim(R)$; the choice affects the numerical form of trace identities but not the structural discussion.} (see e.g.,\ \cite{Nakahara:2003,BaezMuniain:1994}). 

Cyclicity of the trace implies a basic exchange identity for loops based at the same point. Using the composition
convention of relation \eqref{eq:concat_rule},
\begin{equation}
W_R(C_1\circ C_2)=\mathrm{Tr}_R\,\big(U_R(C_1)\,U_R(C_2)\big)
=\mathrm{Tr}_R\,\big(U_R(C_2)\,U_R(C_1)\big)=W_R(C_2\circ C_1).
\end{equation}
For unitary representations one also has orientation reversal
\begin{equation}
W_R(C^{-1})=\mathrm{Tr}_R\,U_R(C)^{-1}=\overline{W_R(C)}.
\end{equation}
In contrast to the holonomy level concatenation rule, traced Wilson loops are not multiplicative in general:
\begin{equation}
W_R(C_1\circ C_2)\neq W_R(C_1)\,W_R(C_2).
\end{equation}
Commutativity of the holonomies is not sufficient to obtain multiplicativity, since even for commuting matrices
$\mathrm{Tr}(AB)$ need not equal $\mathrm{Tr}(A)\mathrm{Tr}(B)$. Multiplicativity is recovered in a one dimensional
representation, as in the $U(1)$ example below, and in certain other special cases. Accordingly, admissibility of traced loop data is expressed not by a general product rule but by nontrivial
constraints, including Cayley-Hamilton/Mandelstam type trace identities, positivity or inequality conditions, and
regularity requirements (see, e.g.,\ \cite{Loll:1993,Polyakov:1987gfs,Makeenko:2002mcgt}).

\medskip
\noindent

\paragraph*{Example: $G=U(1)$.} Consider the standard one dimension representation of $U(1)$. Since the group is abelian, path ordering is
unnecessary, and for a path $\gamma_{x,y}$,
\begin{equation}
U(\gamma_{x,y})=\exp\,\left(i\int_{\gamma_{x,y}} A\right).
\end{equation}
Under a gauge transformation $A\mapsto A+d\alpha$ one finds
\begin{equation}
U(\gamma_{x,y})\mapsto e^{i\alpha(x)}\,U(\gamma_{x,y})\,e^{-i\alpha(y)},
\end{equation}
so open lines are endpoint covariant, while for a closed loop $C$ the phase cancels and
\begin{equation}
U(C)=\exp\,\left(i\oint_C A\right)
      =\exp\,\left(i\int_\Sigma F\right),\qquad F=dA,
\end{equation}
for any surface $\Sigma$ with $\partial\Sigma=C$ within a contractible patch. In this one dimension representation, the traced Wilson loop coincides with the holonomy itself. Therefore,
\begin{equation}
W(C_1\circ C_2)=W(C_1)\,W(C_2).
\end{equation}
Thus multiplicativity holds in this abelian one dimension example, whereas for Wilson loops defined by traces of
higher dimensions matrices the trace generally does not preserve the holonomy product. Such this contrast with higher dimensions non-Abelian representations will be made explicit below through the $SU(2)$ and $SU(3)$ trace identities.

\subsection{Mandelstam type trace identities (finite dimensionality constraints)}
\label{ss:mandelstam}

In a fixed finite dimension representation $R$, a traced Wilson loop is the character of the loop holonomy, Equation \eqref{eq:wilson_loop_def}, and is therefore a class function: it depends only on the conjugacy class of $U(C)$.
Consequently, Wilson loop data cannot be prescribed arbitrarily as a function of loops: even before considering relations among different loops, the value assigned to a loop can depend on its holonomy only through this conjugacy invariant information.

A further kinematical restriction comes from finite dimensionality. If $\dim R=d_R$, then every matrix
$U_R(C)$ satisfies its degree $d_R$ characteristic equation by the Cayley-Hamilton theorem
(see Appendix~\ref{app:CH}), where it says sufficiently high powers of $U_R(C)$ can be expressed in terms of lower
powers. Taking traces gives polynomial relations among the winding observables
\begin{equation}
W_R(C^n)=\mathrm{Tr}_R\,\big(U_R(C)^n\big),
\end{equation}
where $C^n$ denotes traversing the same loop $n$ times. The same finite dimension matrix identities also produce
relations among traces of products of holonomies associated with loops composed at a common base point.

In the loop variable literature, these polynomial trace relations are commonly referred to as Mandelstam identities or Mandelstam constraints; see, e.g.,\ the original path dependent formulation and subsequent loop space and reconstruction analyses
\cite{Mandelstam:1968,Giles:1981reconstruct,Loll:1992,Barrett:1991holonomy}.
Two points are particularly relevant for reconstruction:

\begin{enumerate}[label=(\arabic*)]
\item Loss of information and representation dependence.
Knowing $W_R(C)$ in a single representation does not in general determine $U(C)$, even up to conjugation: it gives
only the value of one character on the conjugacy class of the holonomy.\footnote{Whether one character is sufficient
to distinguish conjugacy classes depends on the group $G$ and the representation $R$. For a compact group, the
characters of all irreducible representations together separate conjugacy classes, but a single character need not
do so.}
Depending on $G$ and $R$, one may therefore need additional independent class functions, such as traces in other
representations or traces of powers of the holonomy, to characterize the conjugacy class more fully.
This is the additional loss of information caused by tracing, beyond the conjugation freedom already present for
based holonomies.

\item Mandelstam constraints are not the whole story.
Even if a proposed loop functional $C\mapsto W_R(C)$ satisfies all polynomial trace identities appropriate to $G$ and $R$, this is generally not sufficient to guarantee that it arises from a smooth connection. The loop data must also satisfy suitable regularity or continuity conditions under smooth loop deformations, together with positivity and inequality constraints on the allowed values of traced holonomies; see, e.g.,\ \cite{Loll:1993ineq}.
\end{enumerate}

\subsection{From traced loop data to a holonomy representation}
\label{ss:loop_to_holonomy}

Earlier we showed that once a holonomy level assignment $\gamma\mapsto U(\gamma)$ is available, the local gauge potential
can be recovered from infinitesimal segments. What remains is the inverse step:

\medskip
\noindent
\textbf{Question.}
Given only traced loop data $W_R(C)$, under what conditions can one construct a compatible based loop holonomy assignment
$C\mapsto U_R(C)$, defined up to simultaneous conjugation at the base point, such that
\begin{equation}
W_R(C)=\mathrm{Tr}_R\,U_R(C)\,?
\end{equation}
For the positivity argument below, we assume that the chosen finite dimension representation $R$ is unitary.
This covers the standard case of compact gauge groups; for nonunitary representations the corresponding positivity
condition would require a modified formulation. The construction first
concerns loops based at one fixed point; once a compatible based loop holonomy assignment is available, the
reference path construction of Section~\ref{sec:recover} can be used to recover open holonomies.

\medskip
\noindent
A convenient way to organize this lifting problem is to encode the loop functional as a positive linear functional on
a $*$-algebra and then apply the GNS construction to obtain a $*$-representation.

\subsubsection{Based loops and the holonomy representation problem}

Fix $x\in M$ and let $\mathcal{L}_x$ denote the piecewise smooth loops based at $x$. Loops that differ only by an orientation preserving reparametrization are regarded as equivalent, and a path segment followed immediately by the same segment in reverse is taken to contribute nothing.
With these conventions, concatenation gives the product, reversal gives the inverse. Thus, concatenation and reversal then define a group structure on $\mathcal{L}_x$, with the constant loop $\mathrm{id}_x$ as the identity. A holonomy level description is, abstractly, a map
\begin{equation}
\rho:\mathcal{L}_x \longrightarrow U(\mathcal{H})
\end{equation}
into unitary operators on some Hilbert space $\mathcal{H}$ such that
$\rho(C_1\circ C_2)=\rho(C_1)\rho(C_2)$ and $\rho(C^{-1})=\rho(C)^{-1}$.
The finite dimension gauge theory situation corresponds to
$\mathcal{H}\cong\mathbb{C}^{d_R}$, with $\rho(C)$ represented by a $d_R\times d_R$ unitary matrix in the chosen representation $R$.

\subsubsection{The loop $*$-algebra and the induced functional}
Consider the complex group algebra of based loops,
\begin{equation}
\mathcal{A}_x := \mathbb{C}[\mathcal{L}_x]
=\left\{\sum_{i=1}^n \alpha_i\, C_i \;\middle|\; n<\infty,\ \alpha_i\in\mathbb{C},\ C_i\in \mathcal{L}_x\right\},
\end{equation}
with multiplication induced by concatenation and unit given by the constant loop (denoted $\mathbbm{1}$ or $\mathrm{id}_x$ interchangably.).
Orientation reversal induces a natural involution: set $C^*:=C^{-1}$ for $C\in\mathcal{L}_x$ and extend antilinearly,
\begin{equation}
\left(\sum_i \alpha_i C_i\right)^* := \sum_i \overline{\alpha_i}\, C_i^{-1}.
\end{equation}
Given traced loop data in a fixed representation $R$, define a linear functional
$\omega_R:\mathcal{A}_x\to\mathbb{C}$ by linear extension,
\begin{equation}
\omega_R\,\left(\sum_i \alpha_i C_i\right) := \sum_i \alpha_i\, W_R(C_i).
\end{equation}
Because $W_R$ is obtained by taking a trace, the functional must satisfy
\begin{equation}\label{eq:tracial_functional}
\omega_R(ab)=\omega_R(ba),
\qquad a,b\in\mathcal{A}_x.
\end{equation}
This is the algebraic form of the exchange identity for based loops discussed above. With the unnormalized trace
convention used in this review,
\begin{equation}
\omega_R(\mathrm{id}_x) = W_R(\mathrm{id}_x) = d_R.
\end{equation}
A further necessary condition for $W_R$ to arise from a unitary holonomy representation is positivity:
\begin{equation}\label{eq:positive_functional}
\omega_R(a^*a)\ \ge\ 0
\qquad\text{for all } a\in\mathcal{A}_x.
\end{equation}
Equivalently, for any finite set of loops $C_1,\dots,C_n$ and coefficients $\alpha_1,\dots,\alpha_n$ one must have
\begin{equation}
\sum_{i,j=1}^n \overline{\alpha_i}\alpha_j\; W_R(C_i^{-1}\circ C_j)\ \ge\ 0.
\end{equation}
Thus, for every finite collection of based loops, the matrix
\begin{equation}
\left[W_R(C_i^{-1}\circ C_j)\right]_{i,j=1}^{n}
\end{equation}
is positive semidefinite. A loop functional satisfying this condition is said to be of positive type.

\subsubsection{GNS construction: producing a $*$-representation from $\omega_R$}

Up to this stage, the traced loop data assigns one complex number $W_R(C)$ to each based loop $C$.
To move toward a holonomy level description, we need a way to represent loops by operators while reproducing these
numbers. This is the role of the Gel'fand-Naimark-Segal (GNS) construction. Starting from the positive functional
$\omega_R$ on the loop $*$-algebra $\mathcal{A}_x$, it constructs a Hilbert space $\mathcal{H}_R$, a
$*$-representation $\pi_R$ of $\mathcal{A}_x$, and a distinguished vector $\Omega$ such that the original functional is recovered as a matrix element of the representation.

Therefore, GNS provides a first step from scalar traced loop data to an operator valued representation of the loop algebra. However, it does not, by itself, recover the original $d_R$-dimension holonomy matrices, nor does it yet realize $W_R(C)$ as the trace of $\pi_R(C)$. Those additional requirements involve finite dimensionality and character constraints, which are discussed below. We now give the construction explicitly (see, e.g.,\
\cite{BratteliRobinson,KadisonRingrose}). 


\medskip

Given a positive functional $\omega_R$ on a $*$-algebra, GNS construction produces a canonical
cyclic $*$-representation realizing $\omega_R$ as a vector state (see e.g.,\ \cite{BratteliRobinson,KadisonRingrose}).
Define the null space
\begin{equation}
\mathcal{N}:=\{a\in\mathcal{A}_x \mid \omega_R(a^*a)=0\}.
\end{equation}
The elements of $\mathcal{N}$ have zero length with respect to the positive form defined by $\omega_R$.
We therefore identify two elements $a,b\in\mathcal{A}_x$ whenever $a-b\in\mathcal{N}$, and denote the equivalence
class of $a$ by $[a]$. The Cauchy-Schwarz inequality following from positivity ensures that
\begin{equation}
\langle [a],[b]\rangle := \omega_R(a^*b)
\end{equation}
is a well-defined inner product on the quotient $\mathcal{A}_x/\mathcal{N}$. Let $\mathcal{H}_R$ be the Hilbert space completion of $\mathcal{A}_x/\mathcal{N}$, and define
$\pi_R:\mathcal{A}_x\to \mathcal{B}(\mathcal{H}_R)$ by left multiplication,
\begin{equation}
\pi_R(a)\,[b] := [ab].
\end{equation}
For the loop algebra considered here, the null space $\mathcal{N}$ is preserved under left multiplication, so this definition does not depend on the representative chosen for $[b]$. Then $\pi_R$ is a $*$-representation, i.e.\ $\pi_R(a^*)=\pi_R(a)^\dagger$. Now let
\begin{equation}
\Omega:=[\mathbbm{1}].
\end{equation}
This vector is cyclic, meaning that vectors of the form $\pi_R(a)\Omega$, with $a\in\mathcal{A}_x$, span a dense
subspace of $\mathcal{H}_R$. The GNS reconstruction identity is
\begin{equation}\label{eq:gns_state}
\omega_R(a)=\langle \Omega,\pi_R(a)\Omega\rangle,
\qquad a\in\mathcal{A}_x.
\end{equation}
With the unnormalized trace convention used in this review,
\begin{equation}
\|\Omega\|^2=\omega_R(\mathbbm{1})=d_R.
\end{equation}
For a loop $C\in\mathcal{L}_x$, the operator $\pi_R(C)$ is unitary because
$C^*C=CC^*=\mathbbm{1}$, and
\begin{equation}\label{eq:loop-matrix-element}
W_R(C)=\langle\Omega,\pi_R(C)\Omega\rangle.
\end{equation}
Thus the GNS construction reproduces the traced loop functional as a matrix element of an operator representation. At this stage, this matrix element should not be confused with the finite dimension trace
$\mathrm{Tr}_R\,U_R(C)$ from which the original data are assumed to arise.

\paragraph*{Kernel/quotient interpretation:}
The quotient has a direct meaning: it identifies algebra elements whose difference has zero length according to the
loop functional. In particular, for loops $C,C'\in\mathcal{L}_x$, viewed as elements of $\mathcal{A}_x$, one has
$[C]=[C']$ precisely when
\begin{equation}\label{eq:equivN}
\omega_R\,\big((C-C')^*(C-C')\big)=0.
\end{equation}
By the Cauchy-Schwarz inequality, an element with zero length has zero inner product with every element of the GNS space. Therefore Equation \eqref{eq:equivN} is equivalent to
\begin{equation}
W_R(C^{-1}\circ Z)=W_R(C'^{-1}\circ Z)\qquad
\text{for every }Z\in\mathcal{L}_x.
\end{equation}
Thus two loops represent the same vector in the GNS space exactly when no further concatenation with a based loop
$Z$ distinguishes them through the given traced loop functional.


\subsubsection{Finite dimensionality and character constraints}


The GNS representation $\pi_R$ constructed above need not be finite dimensional. Moreover, relation
\eqref{eq:gns_state} realizes the loop functional as a matrix element, relation \eqref{eq:loop-matrix-element}, rather than as the trace of a $d_R$-dimension matrix representation (a brief reminder of characters and reducibility is given in Appendix~\ref{app:repthm}). Remember that recovering matrices $U_R(C)$ satisfying $W_R(C)=\mathrm{Tr}_R\,U_R(C)$, and therefore requires an additional step: one must determine whether the prescribed loop functional is the character of a $d_R$- dimension unitary representation of the based loop group.

At the algebraic level, necessary conditions include the normalization $W_R(\mathbbm{1})=d_R$, the trace or exchange property, positivity, and the finite dimension trace identities appropriate to the representation $R$. The latter include the Mandelstam/Cayley-Hamilton type relations collected in Appendix~\ref{app:mandelstam}. These conditions strongly restrict the allowed loop functional, but they do not by themselves give a complete existence theorem in full generality.\footnote{Depending on the gauge group and the representation content, additional representation specific conditions may be required to establish that the functional is the character of a $d_R$-dimension unitary representation.} Regularity and continuity under smooth loop deformations form a further, logically separate requirement. They are
needed to ensure that a resulting holonomy assignment can be interpreted as parallel transport generated by a
smooth connection. When a compatible finite dimension character realization exists, its matrices provide the
required based loop holonomies $U_R(C)$.

\subsubsection{From based loop holonomy to $A_\mu(x)$}

Once the traced loop data has been lifted to a compatible finite dimension based loop holonomy assignment, the
construction of Section~\ref{sec:recover} can be used, together with a choice of reference holonomies, to extend it
to open paths. If the resulting open path holonomies have the required smooth dependence on short segments, then $U(S_{x+\epsilon,x})
= \mathbbm{1}+i\,\epsilon^\mu A_\mu(x)+\mathcal{O}(\epsilon^2)$, and hence $A_\mu(x)$ by Equation \eqref{eq:A_component_from_U} as derived in Section~\ref{subsec:infinitesimal_extraction}. 

Thus, after the traced data has been lifted to sufficiently
regular open path holonomies, the local gauge potential is recovered as the infinitesimal generator of parallel
transport.





\medskip
\noindent
The reconstruction program can therefore be organized into four steps:

\begin{enumerate}
    \item traced loop data satisfying the trace and positivity conditions define a positive tracial functional on the
loop $*$-algebra;

\item the GNS construction turns this functional into a cyclic $*$-representation that reproduces the loop data as
matrix elements;

\item a finite dimension character realization, when it exists, produces matrix valued based loop holonomies
$U_R(C)$ satisfying $W_R(C)=\mathrm{Tr}_R\,U_R(C)$;

\item extension to sufficiently regular open path holonomies allows the local connection to be recovered from the infinitesimal segment limit.
\end{enumerate}
The third step is not automatic: it is precisely where the finite dimension trace identities and additional
character constraints enter.



\subsection{Worked examples of trace identities}\label{sub:wexam}

To make the finite dimensionality constraints concrete, we record two standard examples in the \emph{defining}
(fundamental) representation of $G=SU(2)$ and $G=SU(3)$.
Throughout this subsection, $U(C)$ denotes the holonomy matrix in that representation and
\begin{equation}
W(C)=\mathrm{Tr}\,U(C),
\end{equation}
denotes its trace. We write $C^n$ for the loop obtained by traversing the same loop $C$ $n$ times, so that
$U(C^n)=U(C)^n$. When loops are composed as $C_1\circ C_2$ we assume they share a common base point.

\subsubsection{Example: $G=SU(2)$}
For any $2\times2$ matrix $U\in SU(2)$, Cayley-Hamilton implies
\begin{equation}
U^2-(\mathrm{Tr}\,U)\,U+\mathbbm{1}=0,
\end{equation}
since $\det U=1$. Applying this to $U(C)$ and taking the trace gives
\begin{equation}
W(C^2)=\mathrm{Tr}\,\big(U(C)^2\big)=W(C)^2-2.
\end{equation}
A second identity relates products of traces to traces of composed holonomies:
\begin{equation}
W(C_1)\,W(C_2)=W(C_1\circ C_2)+W(C_1\circ C_2^{-1}).
\end{equation}
In particular, in the defining representation of $SU(2)$ the conjugacy class of $U(C)$ is determined by $W(C)$
(equivalently by the eigenvalue angle), so a single traced loop already captures the conjugacy class information for each loop.

\subsubsection{Example: $G=SU(3)$}
For $U\in SU(3)$, Cayley-Hamilton yields
\begin{equation}
U^3-(\mathrm{Tr}\,U)\,U^2+\frac12\left[(\mathrm{Tr}\,U)^2-\mathrm{Tr}(U^2)\right]U-\mathbbm{1}=0,
\end{equation}
and taking the trace gives
\begin{equation}
\mathrm{Tr}(U^3)=\frac32\,(\mathrm{Tr}\,U)\,\mathrm{Tr}(U^2)-\frac12\,(\mathrm{Tr}\,U)^3+3.
\end{equation}
For a loop $C$ this becomes
\begin{equation}
W(C^3)=\frac32\,W(C)\,W(C^2)-\frac12\,W(C)^3+3.
\end{equation}
For $U\in SU(3)$, unitarity and $\det U=1$ imply
\begin{equation}
\mathrm{Tr}(U^{-1})=\overline{\mathrm{Tr}(U)}, \qquad \text{and} \qquad \mathrm{Tr}(U^2)= \big(\mathrm{Tr}\,U\big)^2-2\,\overline{\mathrm{Tr}(U)}.
\end{equation}
Thus the full complex trace in the defining representation determines the characteristic polynomial of $U$ and
hence its conjugacy class. In loop notation,
\begin{equation}
W(C^2)
=
W(C)^2
-
2\,W(C^{-1}),
\end{equation}
and the cubic identity may equivalently be written as
\begin{equation}
W(C^3)
=
W(C)^3
-
3\,W(C)\,W(C^{-1})
+
3.
\end{equation}
The $SU(3)$ example therefore illustrates how finite dimensionality relates different windings of the same loop.
For groups and representations in which a single character does not contain enough independent conjugacy class
information (for example, the defining representation of $SU(N)$ with $N\geq4$) additional class functions are
needed. Detailed derivations are given in Appendix~\ref{app:SU2SU3}.


\subsection{Other reconstruction viewpoints and related literature}
\label{sec:other_reconstruction}

The phrase "reconstructing a gauge field from Wilson loops" is used in the literature in several distinct senses.
For orientation, it is useful to separate: (a) local geometric reconstruction from infinitesimal loop holonomies, (b) loop space formulations in which loop observables are treated as primary variables, and (c) lattice inverse problems that infer effective continuum descriptions from ensembles of loop observables. The present review focuses mainly on (a) and on kinematical existence/uniqueness questions for holonomy/character data.

\begin{enumerate}[label=(\alph*)]
\item Infinitesimal loops, curvature, and local reconstruction. On a contractible patch, holonomy around sufficiently small loops determines the curvature to leading order; schematically, for a loop of linear size $\epsilon$ and enclosed area $\Sigma^{\mu\nu}=\mathcal{O}(\epsilon^2)$ one has
\begin{equation}
U(C_{\epsilon})=\mathbbm{1}+ i\,F_{\mu\nu}(x)\,\Sigma^{\mu\nu}+\mathcal{O}(\epsilon^3),
\end{equation}
in a chosen representation (compare the explicit rectangular loop expansion in \eqref{eq:small_loop_expansion}).
This is the basic local mechanism behind extracting pointwise information from nonlocal parallel transport and is closely related to the general differential geometric relation between curvature and holonomy (e.g.,\ the Ambrose -Singer theorem; see e.g.,\ \cite{KobayashiNomizu:1963fdg1}).
In the abelian case Stokes' theorem relates $\oint A$ directly to $\int F$; in the non-Abelian case path ordering obstructs
a naive Stokes formula, but one can still represent holonomy in terms of $F$ using surface ordered (non-Abelian Stokes) constructions at the price of introducing surface ordering and parallel transports to a reference point, see e.g.,\
\cite{Fishbane:1981,Diosi:1983,2001JETP...92..905D,HirayamaUeno:2000nast}.

\item Loop space variables and loop equations.
A different use of reconstruction appears in loop space approaches that attempt to reformulate gauge theory in terms of loop observables. In large $N$ and related settings, the Makeenko-Migdal loop equations (loop Schwinger-Dyson equations) provide functional relations among Wilson loop expectation values $\langle W_R(C)\rangle$, rather than pointwise reconstruction
of a specific classical connection. These relations act as dynamical/consistency constraints on loop data, but they should be distinguished from the kinematical reconstruction problem studied here (existence and uniqueness up to gauge of a connection compatible with given holonomy or character data), see e.g.,\ \cite{MakeenkoMigdal:1981loops,Makeenko:2002mcgt,Polyakov:1987gfs}.

\item Lattice gauge theory and inverse problems.
On the lattice, the fundamental variables are link matrices (discrete parallel transports), and Wilson loops are traces of ordered link products. Smoothing procedures such as cooling/smearing and gauge covariant flows are widely used to suppress ultraviolet noise and to define continuum like effective fields; these methods act on gauge configurations, while Wilson loops serve as diagnostics, see e.g.,\ \cite{Wilson:1974sk,Rothe:2005lattice,Albanese:1987,MorningstarPeardon:2004,Luscher:2010}.
Inverse Monte Carlo/demon methods and Monte Carlo renormalization group approaches address a different inverse problem: from measured loop observables one infers an effective action (couplings) reproducing those observables at the level of an ensemble, rather than reconstructing a unique underlying configuration $A_\mu(x)$, see e.g.,\ \cite{Creutz:1983,Swendsen:1984}.

\end{enumerate}
Finally, it is essential to distinguish traced from untraced data. A smooth assignment $\gamma\mapsto U(\gamma)$ for all (open) paths satisfying the holonomy/parallel transport axioms determines a compatible connection locally up to gauge under mild regularity assumptions, see e.g.,\ \cite{Giles:1981reconstruct,Barrett:1991holonomy}.
In contrast, traced Wilson loops in a single fixed representation provide only character data and therefore determine at most conjugacy class information subject to trace identities (Mandelstam/Cayley-Hamilton) and additional admissibility conditions; lifting from traced data to full holonomy data typically requires additional representation content and/or additional structure.

\subsection{Renormalization and products of Wilson loops}
\label{subsec:wilson_loop_renormalization}


The reconstruction statements developed above concern classical holonomies. In quantum gauge theory, a Wilson loop
is a nonlocal composite operator: it is constructed from the gauge field along an entire contour rather than at a
single spacetime point. Its expectation value can therefore contain short distance (ultraviolet) divergences even
after the usual renormalization of the fields and coupling.

In fact, the additional renormalization depends on the geometry of the contour. For a smooth, non self-intersecting loop, a cutoff regularization may produce a divergence proportional to the total length of the contour. This is commonly called a perimeter divergence and actually can be absorbed into a multiplicative factor associated with the loop. However, if the contour contains a cusp - a point where its tangent changes discontinuously - then an additional logarithmic divergence appears. Therefore, the corresponding renormalization factor depends on the cusp angle, and its dependence on the renormalization scale is described by the cusp anomalous dimension
\cite{Polyakov:1980rings,KorchemskyRadyushkin:1987}.

Furthermore, a self-intersection requires a different treatment. At an intersection, the same incoming and outgoing path segments can be joined in more than one way. The Wilson loops associated with these different connections generally mix under renormalization, so they must be renormalized together rather than by assigning a single multiplicative factor to one loop
\cite{BrandtNeriSato:1981}. This distinction becomes important especially for products of Wilson loops. If two smooth contours are separated, the renormalization factors associated with the individual loops can be treated independently. While if the contours intersect, or share part of a path, additional short distance singularities may appear, and therefore the product can
mix with other loop operators. A concrete example is the cyclic Wilson loop at finite temperature, which mixes under
renormalization with the correlator of two Polyakov loops
\cite{BerweinBrambillaVairo:2013}.

Nevertheless, the classical composition and trace identities discussed in this review remain exact identities for holonomy matrices. They should not, however, be applied directly to renormalized quantum Wilson loop operators or their expectation values without considering the contour dependent renormalization factors and possible operator
mixing. Therefore, in this cases, the contour geometry and renormalization prescription and the set of operators that mix must be specified as part of the problem.

\section{Conclusions and outlook}\label{sec:sum}


This review developed Wilson lines and Wilson loops from the viewpoint of parallel transport in continuum gauge theory.
The starting point was the fact that local gauge transformations prevent a direct comparison of fields at
separated points. A Wilson line provides the required gauge covariant transport along an oriented path, while for a
closed path the holonomy transforms by conjugation and its trace defines a gauge invariant Wilson loop. The distinction
between the untraced holonomy $U(C)$ and the traced Wilson loop $W_R(C)=\mathrm{Tr}_R\,U_R(C)$ has been the organizing theme of the review, since it determines what information is
available in reconstruction problems.

Starting from the transport equation, we obtained the Wilson line as a path ordered exponential and collected its
basic structural properties: identity, composition under path concatenation, inversion under path reversal, and
reparametrization invariance. We also showed how finite transport arises as the ordered continuum product of
infinitesimal segment holonomies. This construction explains both the multiplicative nature of parallel transport and the need for path ordering in a non-Abelian theory.



Local gauge field information is recovered through two complementary limits. The short segment expansion of $U(S_{x+\epsilon,x})$ identifies the connection as the infinitesimal generator of parallel transport. Fixed endpoint path deformations
insert the curvature $F_{\mu\nu}$ into the Wilson line, while small loop expansions recover the curvature from the
leading area dependence of the holonomy. Short segments and small loops therefore provide the direct link between
local gauge field data and nonlocal transport. We also explained how these local expressions are supplemented by
transition functions when a Wilson line passes between patches on a topologically nontrivial gauge bundle. Base point dependence was treated separately from path deformation. Transporting a displaced loop holonomy back to
a fixed base point produces a gauge covariant expansion involving the adjoint covariant derivative. The short
connecting segments generate the commutator with $A_\mu(x)$, while the change of the chosen loop family is contained in the derivative of its holonomy and can be related to curvature through the path deformation formula. More generally, changing the base point of an untraced loop holonomy acts by conjugation, whereas taking the trace removes this dependence.


The reconstruction problem emphasized in this review can be summarized cleanly at two levels. At the holonomy level, one asks when an assignment
$\gamma\mapsto U(\gamma)$ is compatible with a smooth connection and whether that connection is unique up to gauge. The identity, composition, inversion, reparametrization, and regularity conditions provide the corresponding consistency requirements, and the local connection is then obtained from the infinitesimal segment limit. We also showed that based loop holonomies at a fixed point, together with a choice of reference paths and reference holonomies, determine open path holonomies. Different choices of reference holonomies act through endpoint factors of the same form as a gauge transformation.


At the traced level, the situation is different because Wilson loops provide character data rather than the full holonomy matrices. They are gauge invariant and base point independent, but a single character may fail to
distinguish conjugacy classes and cannot in general recover the relations among several holonomies. Admissible traced
loop data must satisfy finite dimension trace identities, including Cayley-Hamilton/Mandelstam type relations,
as illustrated by the $SU(2)$ and $SU(3)$ examples. They must also satisfy positivity conditions and suitable
regularity under smooth changes of loops. The loop $*$-algebra and GNS construction organize part of this problem: a positive tracial functional produces a cyclic $*$-representation that reproduces the loop functional as matrix elements. Recovering a $d_R$-dimension matrix representation whose character equals $W_R(C)$ is an additional and nonautomatic step, requiring the appropriate finite dimension character constraints. Once sufficiently regular open path holonomies have been obtained, the connection again follows from the infinitesimal segment limit.

Taken together, these results clarify what it can mean (and what it cannot mean) to reconstruct a gauge field from Wilson loops. If one has sufficiently regular holonomy data on paths (or based loop holonomies supplemented by controlled reference transports), then the local connection is recoverable on a patch up to gauge. If one has only traced Wilson loops in a restricted representation set, then one has gauge invariant information but generally not enough to determine holonomies uniquely, and any
reconstruction requires additional representation content and/or supplementary admissibility assumptions. This is not merely a technical point: it delineates precisely which gauge invariant loop observables are, even in principle, sufficient for local field recovery (up to gauge).

The separation between holonomy level and trace level data also points to several natural directions for further
work. Several parts of the reconstruction program already have precise mathematical and physical formulations, including the reconstruction of connections from compatible holonomy data, algebraic reconstruction results for Wilson loop variables in specified settings, and the renormalization of Wilson loops with cusps or intersections. The remaining challenge is therefore not to establish these ingredients independently, but to connect them in forms that are constructive and useful across different gauge groups, representations, and regularity assumptions. One natural direction is to understand how reconstruction behaves when the available loop data are incomplete, discretized, or affected by numerical uncertainty, as in lattice and other computational settings. A related question is how the classical reconstruction problem should be formulated when the input consists of renormalized quantum expectation values rather than classical holonomies. In that case, contour dependent renormalization and possible mixing among loop operators must be included before the data can be compared with a classical holonomy assignment. These questions represent extensions and practical refinements of the established reconstruction framework developed throughout this review.

theory.

\medskip
\noindent
\medskip
\noindent

 This research is supported in part by the University of Waterloo, Natural Sciences and Engineering Research Council of Canada, by the Government of Canada through the Department of Innovation, Science and Economic Development and by the Province of Ontario through the Ministry of Colleges and Universities at Perimeter Institute. 
  The author would like to thank Niayesh Afshordi, Ifomea Akpuogwu, Avery E. Broderick, Paul Ginsparg, Eva Hackmann, Mengqi Lu, Robert Mann, Samir D. Mathur, Daniel Phillip, Ruijing Tang, and Zhiren Wang for useful discussions and comments.
\section*{Data availability}
No datasets were generated or analysed during the current study.


\bibliographystyle{unsrt}
\bibliography{refwl}


\appendix

\section{Derivation of the gauge field transformation law}
\label{app:gauge_trans_A}

Let $\psi(x)$ be a matter field transforming in some representation of the gauge group $G$, and let
$g:M\to G$ be a local gauge transformation acting as $\psi\mapsto \psi^{\,g}=g\,\psi$ (we suppress the representation map).
With the convention
\begin{equation}
D_\mu:=\partial_\mu-iA_\mu,
\end{equation}
gauge covariance of the derivative means
\begin{equation}
(D_\mu\psi)^{\,g}=g\,D_\mu\psi.
\end{equation}
Equivalently, writing $D_\mu^{\,g}:=\partial_\mu-iA_\mu^{\,g}$, we require
\begin{equation}
D_\mu^{\,g}\psi^{\,g}=g\,D_\mu\psi.
\end{equation}
Expanding both sides gives
\begin{equation}
(\partial_\mu-iA_\mu^{\,g})(g\psi)=(\partial_\mu g)\psi+g\,\partial_\mu\psi-iA_\mu^{\,g}g\psi,
\qquad
g(\partial_\mu-iA_\mu)\psi=g\,\partial_\mu\psi-i\,gA_\mu\psi.
\end{equation}
Cancelling the common term $g\,\partial_\mu\psi$ and using that the relation must hold for all $\psi$ yields
\begin{equation}
(\partial_\mu g)-iA_\mu^{\,g}g=-igA_\mu.
\end{equation}
Multiplying on the right by $g^{-1}$ and solving for the transformed gauge potential, we obtain
\begin{align}
A_\mu^{\,g}=gA_\mu g^{-1}-i(\partial_\mu g)\,g^{-1}
=gA_\mu g^{-1}+i\,g\,\partial_\mu(g^{-1}),
\end{align}
where the second form follows from
$(\partial_\mu g)g^{-1}=-g\,\partial_\mu(g^{-1})$.
This is Equation~\eqref{eq:gauge_trans_A}.

Writing the connection as the $\mathfrak{g}$-valued one-form
$A=A_\mu\,dx^\mu$, the same transformation law becomes
\begin{align}
A^{\,g} = gAg^{-1} -i\,(dg)g^{-1} = gAg^{-1} +i\,g\,dg^{-1}.
\end{align}
If instead one uses
$D_\mu=\partial_\mu-i g_{\mathrm{YM}}A_\mu$, then
\begin{align}
A_\mu^{\,g}&=gA_\mu g^{-1}-\frac{i}{g_{\mathrm{YM}}}(\partial_\mu g)g^{-1}=gA_\mu g^{-1}+\frac{i}{g_{\mathrm{YM}}}g\,\partial_\mu(g^{-1}).
\end{align}


\section{Cayley-Hamilton, traces, and the origin of polynomial identities}
\label{app:CH}

Let $V$ be a complex vector space of dimension $d$ and let $M\in \mathrm{End}(V)$.
The characteristic polynomial of $M$ is
\begin{equation}
p_M(\lambda)\equiv \det(\lambda \mathbbm{1}-M)
=\lambda^{d}-s_1(M)\lambda^{d-1}+s_2(M)\lambda^{d-2}-\cdots+(-1)^d s_d(M),
\end{equation}
where $s_k(M)$ are the elementary symmetric polynomials in the eigenvalues of $M$.
In particular $s_1(M)=\mathrm{Tr}(M)$ and $s_d(M)=\det(M)$. By the Cayley-Hamilton theorem, $M$ satisfies its own characteristic equation,
\begin{equation}
p_M(M)=0.
\end{equation}
Define the power traces by
\begin{equation}
t_n(M):=\mathrm{Tr}(M^n), \qquad t_0(M)=\mathrm{Tr}(\mathbbm{1})=d.
\end{equation}
Multiplying the Cayley-Hamilton identity by $M^{n-d}$ for $n\ge d$ and taking the trace gives

\begin{equation}\label{eq:CH_trace_recurrence}
t_n(M)-s_1(M)\,t_{n-1}(M)+s_2(M)\,t_{n-2}(M)-\cdots+(-1)^d s_d(M)\,t_{n-d}(M)=0,
\qquad n\ge d,
\end{equation}
The coefficients $s_k(M)$ can themselves be expressed in terms of the power traces
$t_1(M),\ldots,t_k(M)$ through the Newton identities. Consequently, once the first $d$ power traces are fixed,
all higher traces $t_n(M)$ with $n>d$ are determined polynomially. The sequence
$\{\mathrm{Tr}(M^n)\}$ therefore cannot be specified independently for all $n$.

When $M=U_R(C)$ is a holonomy matrix in a $d_R$-dimension representation,
\eqref{eq:CH_trace_recurrence} gives the trace identities relating different windings of the same loop,
$W_R(C^n)=\mathrm{Tr}_R(U_R(C)^n)$. More general Mandelstam identities involving several loops arise from the
corresponding finite dimension identities applied to products or combinations of holonomy matrices; see
Appendix~\ref{app:mandelstam}.

\medskip

A related matrix identity, useful when working with exponentials of Lie-algebra elements, is
\begin{equation}\label{eq:det_exp_trace}
\det(e^{X})=e^{\mathrm{Tr}(X)}.
\end{equation}
It holds for every finite dimension matrix $X$ (for example, by using its Jordan normal form). In particular,
if $\mathrm{Tr}(X)=0$, then $\det(e^X)=1$. This explains why exponentials of traceless generators have unit
determinant, as required in the defining representation of $SU(N)$.

\section{Mandelstam constraints and finite dimension trace identities}
\label{app:mandelstam}

Let fix a finite dimension representation $R$ of $G$ with $\dim R=d_R$, and write
\begin{equation}
W_R(C)=\mathrm{Tr}_R\,U_R(C).
\end{equation}
Since $U_R(C)\in\mathrm{End}(R)$ satisfies its degree $d_R$ characteristic polynomial, the
Cayley-Hamilton theorem (Appendix~\ref{app:CH}) implies that every power $U_R(C)^n$ with $n\ge d_R$ can be
expressed as a linear combination of
\begin{equation}
\mathbbm{1},U_R(C),\ldots,U_R(C)^{d_R-1},
\end{equation}
with coefficients determined by the characteristic polynomial. Taking traces therefore gives polynomial relations among the winding observables
\begin{equation}
W_R(C^n)=\mathrm{Tr}_R\,\big(U_R(C)^n\big),
\end{equation}
where $C^n$ denotes traversing the same loop $C$ $n$ times.
More generally, finite dimension matrix identities applied to products of holonomy matrices give relations among traces associated with loops composed at a common base point.

In the loop variable literature, these relations are commonly called Mandelstam identities or
Mandelstam constraints. They express the fact that not every assignment $C\mapsto W_R(C)$ can arise from traces of finite dimension holonomy matrices.
For discussions of Mandelstam identities, their relation to finite dimension matrix identities, and their role in
loop-space reconstruction, see, e.g.,\
\cite{Mandelstam:1968,Loll:1992,Loll:1993ineq,BerensteinUrrutia:1994,Giles:1981reconstruct,Barrett:1991holonomy}.

\subsection{A canonical permutation form in the fundamental of $SU(N)$}

In the defining (fundamental) representation of $SU(N)$, one convenient way to package a family of trace identities is via
the vanishing of complete antisymmetrization on $(N+1)$ fundamental indices. This vanishing follows from the fact that in an $N$ dimensions representation one cannot antisymmetrize nontrivially over $N+1$ indices. Translated into trace language, this yields
identities that can be written as alternating sums over permutations.

Concretely, given $(N+1)$ loops $C_1,\dots,C_{N+1}$ based at the same point, write $U_i:=U(C_i)$ (in the fundamental of $SU(N)$).
For each $\pi\in S_{N+1}$ with cycle decomposition
\begin{equation}
\pi = (i_1\,i_2\cdots i_{r_1})\,(j_1\,j_2\cdots j_{r_2})\cdots,
\end{equation}
define
\begin{equation}
  W_\pi(C_1,\dots,C_{N+1}):=\mathrm{Tr}\,\big(U_{i_1}\cdots U_{i_{r_1}}\big)\;
\mathrm{Tr}\,\big(U_{j_1}\cdots U_{j_{r_2}}\big)\cdots .
\end{equation}
The antisymmetrization constraint then gives the trace identity

\begin{equation}\label{eq:mandelstam_perm_form}
  0=\sum_{\pi\in S_{N+1}} (-1)^{\pi}\, W_\pi(C_1,\dots,C_{N+1}).
\end{equation}
Here $(-1)^\pi$ is the parity of $\pi$. Equation~\eqref{eq:mandelstam_perm_form} is the standard
fundamental trace identity for $N\times N$ matrices: each cycle of the permutation gives one trace, and the
alternating sum over all permutations vanishes
\cite{BerensteinUrrutia:1994}.
Although it is written here for holonomies in the defining representation of $SU(N)$, the identity itself follows
from the matrix dimension; the additional condition $\det U_i=1$ gives further group specific relations.

\subsection{Reconstruction relevance}
\label{app:mandelstam_reconstruction}

For reconstruction questions, the key points are:

\begin{enumerate}[label=(\arabic*)]

\item Necessity.
Any candidate traced loop functional $C\mapsto W_R(C)$ must satisfy the finite dimension trace identities
appropriate to the chosen representation. If these Cayley-Hamilton/Mandelstam relations fail, the proposed data
cannot arise as traced holonomy data in that representation.

\item Insufficiency by itself. Satisfying polynomial trace identities is generally not enough: admissible traced data must also
obey positivity/inequality constraints and suitable regularity/continuity conditions (cf.\ the discussion in
Section~\ref{ss:mandelstam} and \cite{Loll:1993ineq}).

\item Loss of information.
Traced loop data is invariant under conjugation and therefore cannot select a particular matrix representative of a holonomy. Moreover, in a fixed representation a single character may not determine even the full conjugacy class.
Recovering based loop holonomies may therefore require additional representation data or other compatible
holonomy level information. Once based loop holonomies have been obtained, the reference holonomies introduced in
Section~\ref{sec:recover} are needed to extend the reconstruction to open paths
\cite{Giles:1981reconstruct,Barrett:1991holonomy}.
\end{enumerate}





\section{Derivations for the $SU(2)$ and $SU(3)$ examples used in the main text}
\label{app:SU2SU3}

This appendix derives the explicit trace identities quoted in Subsection~\ref{sub:wexam}. Throughout, we work in the defining (fundamental) representation.

\subsection{$SU(2)$}

Let $U\in SU(2)$ in the defining $2$ dimensions representation. Cayley-Hamilton gives
\begin{equation}\label{eq:SU2_CH}
U^2-(\mathrm{Tr}\,U)\,U+(\det U)\,\mathbbm{1}=0.
\end{equation}
Since $\det U=1$ in $SU(2)$, this reduces to
\begin{equation}
U^2-(\mathrm{Tr}\,U)\,U+\mathbbm{1}=0.
\end{equation}
Taking the trace yields
\begin{equation}\label{eq:SU2_trace_square}
\mathrm{Tr}(U^2)=(\mathrm{Tr}\,U)^2-2,
\end{equation}
using $\mathrm{Tr}(\mathbbm{1})=2$. Applied to a loop holonomy $U(C)$, this is exactly
$W(C^2)=W(C)^2-2$ with $W(C)=\mathrm{Tr}\,U(C)$.

\medskip
The second identity used in the main text,
\begin{equation}\label{eq:SU2_trace_product}
W(C_1)\,W(C_2)=W(C_1\circ C_2)+W(C_1\circ C_2^{-1}),
\end{equation}
is an immediate consequence of Equation \eqref{eq:SU2_CH}. Indeed, Equation \eqref{eq:SU2_CH} applied to $B\in SU(2)$ implies
\begin{equation}\label{eq:SU2_inverse_from_trace}
B^{-1}=(\mathrm{Tr}\,B)\,\mathbbm{1}-B.
\end{equation}
Multiplying Equation \eqref{eq:SU2_inverse_from_trace} on the left by $A$ and taking the trace gives
\begin{equation}
\mathrm{Tr}(AB^{-1})=(\mathrm{Tr}\,A)(\mathrm{Tr}\,B)-\mathrm{Tr}(AB), \qquad \Longrightarrow \qquad (\mathrm{Tr}\,A)(\mathrm{Tr}\,B)=\mathrm{Tr}(AB)+\mathrm{Tr}(AB^{-1}).
\end{equation}
Setting $A=U(C_1)$ and $B=U(C_2)$ (with $C_1,C_2$ based at a common point) yields \eqref{eq:SU2_trace_product}.
\subsection{$SU(3)$}

Let $U\in SU(3)$ in the defining $3$ dimensions representation. Cayley-Hamilton implies
\begin{equation}\label{eq:SU3_CH}
U^3-(\mathrm{Tr}\,U)\,U^2+\frac12\Big((\mathrm{Tr}\,U)^2-\mathrm{Tr}(U^2)\Big)\,U-(\det U)\,\mathbbm{1}=0.
\end{equation}
With $\det U=1$, taking the trace of Equation \eqref{eq:SU3_CH} gives
\begin{equation}\label{eq:SU3_trace_cube}
\mathrm{Tr}(U^3)=\frac32\,(\mathrm{Tr}\,U)\,\mathrm{Tr}(U^2)-\frac12\,(\mathrm{Tr}\,U)^3+3,
\end{equation}
using $\mathrm{Tr}(\mathbbm{1})=3$. For a loop $C$ this yields
\begin{equation}
W(C^3)=\frac32\,W(C)\,W(C^2)-\frac12\,W(C)^3+3,
\end{equation}
as stated in the main text. 
For $U\in SU(3)$, the second elementary symmetric polynomial in the eigenvalues can also be written as
$\mathrm{Tr}(U^{-1})$. Hence
\begin{equation}\label{eq:SU3_inverse_trace}
\mathrm{Tr}(U^{-1})= \frac12\left[(\mathrm{Tr}\,U)^2-\mathrm{Tr}(U^2)\right].
\end{equation}
Since $U$ is unitary, we have $\mathrm{Tr}(U^{-1})=\overline{\mathrm{Tr}(U)}$, and therefore
\begin{equation}\label{eq:SU3_trace_square}
\mathrm{Tr}(U^2)=(\mathrm{Tr}\,U)^2-2\,\overline{\mathrm{Tr}(U)}.
\end{equation}
For a loop holonomy this becomes
\begin{equation}
W(C^2)=W(C)^2-2\,W(C^{-1}).
\end{equation}
Substituting this result into \eqref{eq:SU3_trace_cube} gives the equivalent cubic identity
\begin{equation}
W(C^3)=W(C)^3-3\,W(C)\,W(C^{-1})+3.
\end{equation}
The full complex value of $W(C)=\mathrm{Tr}\,U(C)$ determines the characteristic polynomial
\begin{equation}
\lambda^3 - W(C)\lambda^2 + \overline{W(C)}\,\lambda-1,
\end{equation}
and therefore determines the conjugacy class of $U(C)$ in $SU(3)$. This differs from higher dimensions cases such
as the defining representation of $SU(N)$ with $N\geq4$, where a single trace generally does not contain enough
independent information to determine the conjugacy class.

\medskip
\noindent

\textbf{Remark.}
For any fixed finite dimension representation $R$ of dimension $d_R$, the Cayley-Hamilton theorem gives
polynomial relations among sufficiently high powers of a holonomy matrix and hence among the corresponding winding
observables $W_R(C^n)$. More general finite dimension matrix identities give relations among traces of products of
different holonomies. This is the kinematical origin of the finite dimensionality constraints discussed in the main
text.
\section{Representations and reducibility}
\label{app:repthm}

Let $H$ be a group. A (finite dimension complex) representation of $H$ is a homomorphism
\begin{equation}
\rho: H\to GL(V),
\end{equation}
where $V$ is a finite dimension complex vector space.

A subspace $W\subset V$ is called $H$-invariant if $\rho(h)W\subset W$ for all $h\in H$. Since each $\rho(h)$ is invertible, this condition is equivalent to $\rho(h)W=W$ for all $h\in H$. The representation is reducible if it has a nontrivial invariant subspace $0\neq W\neq V$, and
irreducible otherwise.

If $W$ is invariant, one can choose a basis adapted to $W$ so that each matrix $\rho(h)$ takes block upper triangular form
\begin{equation}
\rho(h)\;\sim\;
\begin{pmatrix}
\rho_W(h) & * \\
0 & \rho_{V/W}(h)
\end{pmatrix}.
\end{equation}
Here $\rho_W$ is the subrepresentation on $W$, and $\rho_{V/W}$ is the induced representation on the quotient $V/W$.
The off-diagonal block $*$ encodes a possible \emph{extension}: an invariant complement to $W$ need not exist in general.
In other words, reducible does not necessarily mean a direct sum decomposition into invariant subspaces.

\paragraph{Unitary and compact cases: complete reducibility.} If $\rho$ is unitary, there is a Hermitian inner product on $V$ that is preserved by every $\rho(h)$.
The orthogonal complement of an invariant subspace is then also invariant. Repeating this decomposition gives
\begin{equation}
\rho\cong\rho_1\oplus\cdots\oplus\rho_k,
\end{equation}
where the representations $\rho_i$ are irreducible.
For compact $H$, every finite dimension continuous representation can be made unitary by averaging an inner product over the group, so finite dimension representations are completely reducible in this case as well.

\paragraph{Characters and what traces can (and cannot) see.}
The character of $\rho$ is the class function $\chi_\rho(h)=\mathrm{Tr}\,\rho(h)$.
It is constant on conjugacy classes and additive under direct sums: $\chi_{\rho_1\oplus\rho_2}=\chi_{\rho_1}+\chi_{\rho_2}$.
Moreover, even in the block upper triangular situation above,
\begin{equation}
\mathrm{Tr}\,\rho(h)=\mathrm{Tr}\,\rho_W(h)+\mathrm{Tr}\,\rho_{V/W}(h),
\end{equation}
so the character does not detect the off-diagonal extension data. For unitary or compact group representations such extensions split, but for a general representation this provides a direct example of information that is invisible to the character. This is a concrete illustration of the general theme in the main text: passing to traces/characters discards information. In the notation of the main text, $W_R(C)=\mathrm{Tr}_R\,U_R(C)=\chi_R\,\big(U(C)\big)$, so a traced Wilson loop is the character of the representation $R$ evaluated on the loop holonomy.


\end{document}